\documentclass[a4paper,fleqn,usenatbib]{mnras}

\usepackage[T1]{fontenc}
\usepackage{ae,aecompl}

\usepackage{graphicx}	% Including figure files
\usepackage{amsmath}	% Advanced maths commands
\usepackage{amssymb}	% Extra maths symbols
\usepackage{epsfig}
\usepackage[usenames]{color}
\usepackage{pdfpages}
\usepackage{array}
\usepackage{multirow}
\newcolumntype{P}[1]{>{\centering\arraybackslash}p{#1}}
\newcolumntype{L}[1]{>{\arraybackslash}l{#1}}
\newcolumntype{M}[1]{>{\centering\arraybackslash}m{#1}}
\usepackage{tabularx}
\usepackage{enumitem}
\newlist{tabitemize}{itemize}{1}
\setlist[tabitemize]{nosep,
                  topsep= 0pt,
                  partopsep=0pt,
                  leftmargin= *,
                  label=\textbullet,
                  before=\vspace{-0.6\baselineskip},
                  after=\vspace{-\baselineskip}
                  }
\usepackage{booktabs}
\usepackage{adjustbox}
\usepackage{pdflscape}

\newcommand\ahat{\hat{\alpha}}

\title[LOS Effects in the HFF]{Exploring Effects on Magnifications due to Line-of-Sight Galaxies in the Hubble Frontier Fields}

\author[Raney et al.]{
Catie A. Raney,$^{1}$\thanks{E-mail: raney@physics.rutgers.edu}
Charles R. Keeton,$^{1}$
and Sean Brennan$^{1}$
\\
$^{1}$Department of Physics and Astronomy, Rutgers University, 136 Frelinghuysen Road, Piscataway, NJ 08854\\
}

\date{Accepted XXX. Received YYY; in original form ZZZ}

\pubyear{2019}

\begin{document}
\label{firstpage}
\pagerange{\pageref{firstpage}--\pageref{lastpage}}
\maketitle

\begin{abstract}
Cluster lensing has become an important tool in the search for high redshift galaxies through its ability to magnify sources. In order to determine the intrinsic properties of these galaxies, lensing mass models must be constructed to determine the magnification of the images. These models are traditionally two-dimensional, focusing on the mass within the cluster and either ignoring or approximating any contribution from line-of-sight galaxies. In this paper, we present the first full set of three-dimensional mass models of the six Hubble Frontier Fields and use them to test for systematic biases in magnifications due to using the traditional 2D approach. We find that omitting foreground or background galaxies causes image position offsets between 0.1-0.4'', a non-negligible fraction of the typical 0.3-0.7" residuals of current state-of-the-art models. We also find that median image magnifications can shift by up to 6\%, though it is dependent on the field. This can be alleviated in some cases by approximating the mass in the lensing plane, but a 5\% magnification bias still exists in other cases; image position offsets are also improved, but are still present at 0.10''.
\end{abstract}


%%%%%%%%%%%%%%%%%%%%%%%%%%%%%%%%%%%%%%%%%%%%%%%%%%

%%%%%%%%%%%%%%%%% BODY OF PAPER %%%%%%%%%%%%%%%%%%

\section{Introduction} \label{sec:intro}

Galaxy clusters have in the past few decades been recognized as a powerful tool to study the intermediate- and high-redshift Universe. One of the ways in which clusters can be utilized is as cosmic telescopes: due to their high mass density and large area on the sky, they can highly magnify background galaxies. Magnification can produce an image of the source that is both brighter and bigger, i.e. with a larger angular size, than it would have been without lensing. This effect allows intrinsically faint sources to be detected and studied more easily than they would have been otherwise. Lensing is also not impacted by the same selection effects as deep blank field surveys. Studies of galaxy clusters have found 100s$-$1000s of lensed galaxies from the first billion years of cosmic history \citep{kneib2011}, including some recent candidates at $z\sim10-11$ \citep[e.g.][]{zheng2012,coe2013,bouwens2014,salmon2018}. 

The Hubble Frontier Fields program \citep[HFF;][]{lotz2017} sought to use galaxy clusters for exactly this purpose. With very deep, high-resolution photometric data obtained with HST, a large number of lensed image candidates were found. Dedicated spectroscopic surveys confirmed some of these image candidates and found other lensed systems that were not visible in the HST data. This brought the total number of confirmed spectroscopic images in the six fields to almost 400 from $\sim\!130$ sources while even more of the candidates have photometric redshift constraints. These lensed galaxies, specifically those that are high redshift {at $z>6$}, have yielded important constraints on the faint end of the luminosity function \citep[e.g.][]{mcleod2016,bouwens2017}. Results from \citet{oesch2018} suggest that there may not be as many early galaxies as once thought, which could have implications for when reionization occurred. It also increases the importance of those few high redshift galaxies found by lensing surveys. 

In order to place a lensed galaxy on a luminosity function, one must first determine its magnification so that the intrinsic luminosity can be found. This magnification depends on a number of things that lensing mass models must take into account. These models fall into two categories, depending on their modeling technique. Parametric models assign mass to galaxy- and large-scale halos using given density profiles. Free-form, sometimes called nonparametric, models are not confined to certain density profiles and instead use constraints to place mass where needed. Hybrid techniques are, as the name suggests, a combination of the two. They use a free-form approach for the dark matter distribution, but also parametric techniques to assign mass to galaxies. 

Regardless of the technique, the mass models are constrained by the positions and/or fluxes of lensed images. However, due to the fact that clusters are inherently complicated systems, there are a number of systematics and degeneracies one must take into account. These come from a variety of aspects in the modeling process. To name a few: modeling technique (i.e. parametric vs. free-form),  source redshifts, determination of cluster membership, scatter in mass-luminosity scaling relations, number and placement of large-scale halos, and effects due to line-of-sight galaxies and structure.  

Efforts have been made to study some of these \citep[e.g.][]{dalal2005,jullo2010,host2012,daloisio2014} and, more recently, many studies have looked at possible sources of uncertainty, particularly in the HFF.  \citet{priewe2017} analyzed results from the third round of HFF modeling (this work is based on the fourth round) for two fields, Abell 2744 and MACS J0416, finding significant differences and possible degeneracies in the models from various teams. A study by \citet{harvey2016} tested the assumption that parametric modeling techniques make, i.e. that light traces mass, in MACS J0416 and found that the assumption could cause large image position offsets. \citet{meneghetti2017} created data products for two mock clusters, which were then given to multiple HFF modeling teams and the resulting models were compared. \citet{johnson2016} used these mock clusters to look at systematics due to photometric vs. spectroscopic redshifts, while \citet{acebron2017} utilized them to look at the effects of density profile choice and inclusion of cluster substructure. 

With more data and constraints, statistical errors should get smaller. Indeed, errors in predicted image positions in cluster lensing have decreased by $3-5\times$ in the past $\sim\!10$ years. Even so, there are still significant uncertainties in magnifications, likely due to systematic errors which have gained importance as statistical errors have decreased. Since there had not been an in depth study of such systematics when the HFF lens modeling program started, it was designed in such a way to deal with these unknown errors. That is, the program invited and funded multiple teams to model the fields. These teams, which all make different assumptions and use a range of modeling techniques, would in theory produce results which, when combined, include the effects of systematics. Thus the range produced by all models from the various teams would give an estimate of the full error, even if the errors of the individual teams are only statistical and thus underestimated. 

However, this only works for systematics that are included by and treated differently among the teams; conversely, if a source of error or bias is not included by any of the modeling teams, it would not be taken into account. An example is the subject of this paper: line of sight effects due to galaxies. These galaxies have mass that, though small compared to the mass of the cluster, can affect the path travelled by light from a source and thus may still bias results if left out. The mass is sometimes not included, however, due to computational barriers: not all modeling software has the ability to include multiple lens planes representing deflectors at different redshifts. Mass from a line-of-sight (LOS) galaxy can be included in a single plane model, though the mass must be scaled so as to have a reasonable effect on how the light is bent, which provides just an approximation. \citet{caminha2016} explored the possible effects of LOS galaxies on predicted image positions using toy models of Abell S1063; \citet{chirivi2017} studied the effects on both positions and magnifications in MACS J0416. Both found that image positions could be affected by up to 0.3'', and the latter found that magnifications could be biased by $>5\%$.

We seek here to test the importance of LOS effects in all six of the Frontier Fields by determining if they introduce any systematic effect or bias in magnifications and image position offsets. We aim to do this by first creating both 2D and 3D models of each field. From there, we then make mock models of the fields and compare single-plane models with and without LOS galaxies to the multi-plane models. In a companion paper (Raney et al., in prep), we will place these results in context by comparing our models to those released by other teams modeling the six Frontier Fields, and quantifying the systematic differences between the models.

This paper begins in Section 2 with our modeling methodology, including descriptions of multi-plane lens modeling and the different mass components which comprise our models. Section 3 shows the results of our modeling for each of the six Hubble Frontier Fields. In Section 4 we perform a comparison between our different models for each of the six fields, analyzing how the inclusion of LOS galaxies affects the predicted image positions and magnifications. Results are summarized and future work is discussed in Section 5. 

In this paper, we assume a cosmology with $\Omega_m$ = 0.3, $\Lambda$ = 0.7, and $h$ = 0.7. 

\section{Modeling Methodology} \label{sec:method}

\subsection{Overview}

Here we describe the methodology used to obtain the models discussed in this work. The same methods are used for all six fields, though with some small changes due to varying complexities and differences in available data. These field-specific notes will be discussed in detail in the next section, where we will show our modeling results. The models were created during the fourth round (v4) of HFF modeling.

Our modeling is done with \texttt{lensmodel} \citep{keeton2001,mccully2014}, a parametric code with the capability of including multiple lens planes. The mass model is comprised of three components: (1) large-scale halos for the dark matter and/or hot gas; (2) small-scale halos for the cluster members; and, when applicable, (3) small-scale halos for the LOS galaxies. We create three types of models for each of the clusters: 
\begin{itemize}
\item 2D-noLOS models without any LOS galaxies; 
\item 2D models with line-of-sight galaxies included, but with mass scaled to the cluster redshift such that there is only a single lens plane;
\item 3D models with LOS galaxies placed at their respective redshifts using multiple lens planes.
\end{itemize}

The 2D models are publicly available the the HFF website\footnote{https://archive.stsci.edu/prepds/frontier/lensmodels/}. Magnification maps can be constructed through the given convergence $(\kappa)$ and shear $(\gamma)$ maps by using
\begin{equation}
\mu^{-1} = (1-\kappa)^2 - \gamma^2.
\end{equation}
This is true for any source redshift, given that $\kappa$ and $\gamma$ are both rescaled using
\begin{equation}
\kappa,\gamma\propto \frac{D_{ls}}{D_s},
\end{equation}
where $D_{ls}$ and $D_s$ are the angular-diameter distances from the lens to the source and from the observer to the source, respectively. We note that the 3D model is not made publicly available through the website because the parameters cannot be scaled in this way since there is not just one $D_{ls}$ factor, as will be described in the following section. 

\subsection{Multi-plane modeling}

Here we briefly summarize the formalism behind multiple lens plane modeling (more details can be found in \citealt{blandford1986,kovner1987,schneider1992}). For a single plane model, we can write the lens equation as 

\begin{equation}
y = x-\frac{D_{ls}}{D_s}\ahat(x),
\label{eq:simple-lenseq}
\end{equation}
where $x, y$ are the positions in the lens and source planes, respectively, and $\ahat$ is the deflection angle that the light feels from the one lens. It also includes the previously defined $D_{ls}$ and $D_s$ factors. 

In multi-plane formalism, we account for many lens planes which produce multiple deflections. To determine the final observed image positions, however, it is not as simple as just adding up the deflections from the various planes. Consider a system with two lenses at different redshifts where plane 2 is behind plane 1 such that their redshifts are $z_1<z_2$. If we work `backwards', i.e. from the observer towards the source, then we can use Eq. \ref{eq:simple-lenseq} to find the image positions  $x_2$ in plane 2 which will depend on the image positions $x_1$ in plane 1 and the deflection $\ahat_1$ from the lens in plane 1:

\begin{equation}
x_2 = x_1 -\frac{D_{12}}{D_2}\ahat_1(x_1).
\end{equation}

To get the true position of the source $y$, we also must account for the deflection from the second lens. Then the full lens equation from the source plane and including both lens planes will be:

\begin{align}
y&=x_1-\frac{D_{1s}}{D_s}\ahat_1(x_1)-\frac{D_{2s}}{D_s}\ahat_2(x_2)\\
&=x_1-\frac{D_{1s}}{D_s}\ahat_1(x_1)-\frac{D_{2s}}{D_s}\ahat_2\left(x_1-\frac{D_{12}}{D_2}\ahat_1(x_1)\right)
\end{align}

Thus our lens equation becomes a recursive one: image positions in plane $j$ depend on the deflections felt in the planes $i<j$ before it. For a system with $N$ planes, $x_1$ is the observed image position and $x_{N+1}$ is the image position in the source plane, such that as we increase $j$ we are going further from the observer. For this system of many planes, the lens equation in Eq.3 becomes:
\begin{equation}
x_j = x_1- \sum_{i=1}^{j-1}\frac{D_{ij}}{D_j}\ahat_i(x_i).
\label{eq:mp}
\end{equation}

\subsection{Single-plane approximations}

The recursive nature of Eq. \ref{eq:mp} makes models with multiple lens planes more computationally expensive than those with single planes. One may wish to approximate the contribution of these planes instead of actually solving the complicated multi-plane lens equation. This can be done by ignoring the varying image positions between the planes and just summing the deflections from the main cluster lens $l$ and perturbing galaxy $p$:

\begin{equation}
y\approx x-\frac{D_{ls}}{D_s}\left[\ahat_l(x)-\frac{D_{ps}}{D_{ls}}\ahat_p(x)\right].
\label{eq:sp-approx}
\end{equation}
There are actually two approximations going on here: we are both ignoring that deflections from one plane will be felt in another (i.e., the recursive nature previously discussed) and also scaling the deflection from the perturber. This scaling approximation comes from the factor ${D_{ps}}/{D_{ls}}$ in front of $\ahat_p(x)$. Note that one must assume a single source redshift to compute this factor; we use $z_{src}=2$. 

\begin{table*}
\centering
 \begin{tabular}{M{3.5cm} M{2cm} M{2.3cm} M{1.5cm} M{1.5cm} M{2cm} M{2cm}} 
 \hline
   Cluster & Redshift & Spectroscopic (Photometric) Cluster Members  &Large-Scale Halos &LOS Galaxies &Images (Systems) & 2D (3D) RMS (arcsec) \\ 
 \hline\hline
 Abell 2744 & 0.308 & 92 (163) & 3 & 6 & 71 (24) & 0.41 (0.38) \\
 MACS J0416.1-2403 &  0.396 & 146 (61) & 3 & 19 & 95 (35) & 0.53 (0.55) \\ 
 
  MACS J0717.5+3745 & 0.545 & 121 (179) & 4 & 8 & 29 (9) & 0.79 (0.79) \\  

 MACS J1149.4+2223 & 0.543 & 179 (78) & 3 & 7 & 53 (16) & 0.30 (0.31) \\ 

 Abell S1063 & 0.348 & 153 (101) & 2 & 10 & 50 (19) & 0.33 (0.34) \\  

 Abell 370 & 0.375 & 129 (127) & 4 & 15 & 91 (31) & 0.74 (0.73) \\ 
 \hline
\end{tabular}
\caption{Model summary for each cluster. }
\label{tab:infotable}
\end{table*}

\subsection{Model components}
\subsubsection{Image constraints}

The HFF modeling process began with the participating teams ranking all possible lensed image candidates based on whether or not we believed there was enough evidence to determine if the candidate was indeed part of a lensed system. In order for an image to be ranked \textsc{gold}, it must be spectroscopically confirmed and have a certain amount of votes from the modeling teams. \textsc{Silver} images are those that do not have a spectroscopic redshift but which the majority of modeling teams are confident are part of a lensed image system. \textsc{Bronze} images are more tenuous and have less agreement about whether or not they belong to a given system. 

There are some cases in which spectroscopically confirmed images do not have a medal ranking. One common reason is that the images were late additions, and thus not all of the teams voted on them. Another case is when the candidate did not appear in the HST data, causing many teams to give a low rank to an image candidate. When a narrow-band spectrum was released where emission is evident, some teams changed their votes to reflect the new data, while others did not. In the image tables in Appendix \ref{sec:imcat}, we list both such cases under the ranking \textsc{none} with clarifying notes.

We also note that some candidates received a ranking of \textsc{gold} even though we could find no spectra in the literature. It is common for modeling papers to quote a model redshift for an assumed image of a system. Sometimes this represents the redshift which was assumed during the modeling process, e.g. from another image in the proposed system which did have a spectroscopic redshift, or else the redshift that was predicted by the model. This can then be confused for a measured redshift, even if that specific image did not have a spectrum taken. In this work, our tables only give measured, spectroscopic redshifts. For those images that do not have spectroscopic data but we are reasonably certain are part of a system, we use the redshift of the system in our modeling, but leave it out of the tables. 

The \textsc{silver} and \textsc{bronze} candidates are galaxies that show evidence of being lensed images based on morphology, color, photometric redshifts, and/or if image positions match model predictions. Many of the candidates, specifically those in the \textsc{bronze} category, do not show structure, thus matching the images into systems from a single source is difficult. Color is a useful tool, especially with the deep photometric data of the HFF; lensing is an achromatic process, i.e. the color of every image should be the same as the source, so color can be used to rule out candidate images. One can also use the position of the images as a tool: with a lens model, one can predict where the critical curves should be and thus whether or not it is reasonable that a single source could produce a certain image configuration. This should be done iteratively instead of circularly, i.e. one should not use a model constrained with given images to give insight into whether or not those image positions are likely. 

Photometric redshifts are another useful way of determining which images are part of a common system. There is some risk in using photometric redshifts for sources: in the case of a catastrophic failure, model parameters may be forced to change appreciably to fit the incorrect redshift. We note that work by \citet{johnson2016} indicates that photometric redshifts pose a much smaller threat to the fit if used with a strong base of spectroscopic redshifts to anchor the parameters, as is the case for most of the HFF clusters (except for perhaps MACS J0717). 

Our constraints are typically draw from the \textsc{gold} sample; the majority of our image constraints have spectroscopic redshifts and we rely only minimally on photometric redshifts. There are a few cases in which we use a \textsc{gold}-plus sample which draws from the other medal categories so as to boost the number of constraints. 

\subsubsection{Cluster members \& halos}

We use both spectroscopic and photometric data from mostly publicly available catalogs to determine which galaxies are part of the cluster. We make a cut using a magnitude limit of 23.5 mag in the F814W filter and a radial cut at 2' from the field center. We perform 3-sigma clipping on the spectroscopic data around the known cluster redshift and use these confirmed cluster members to find a color-magnitude relation in the data from the F606 and F814W filters. Galaxies that fall within 1-sigma of the relation are included as photometrically-selected cluster members. When this cut is applied to the spectroscopic sample, $\sim$80\% of galaxies added are at the cluster redshift, while the rest are predominantly background galaxies due to their redder colors. We note that while we believe our cluster member selection to be reasonable, it may have some systematic effect on the results, as might the member selections of other teams. We plan to study and quantify these possible effects in future work.

We include our complete galaxy catalogs for the six fields in Appendix \ref{sec:galcat}. These tables include cluster members, both spectroscopically- and photometrically-selected, as well as foreground and background galaxies. They are created by combining multiple publicly available catalogs which are listed as references in the tables. Many of the LOS galaxies in the catalog were not included in our models, but we include them to aid other modelers; some of the galaxies may appear to be cluster members based on color but in reality are not, based on spectroscopic redshifts. 

The member galaxies are modeled as spherical pseudo-Jaffe mass distribution with only Einstein and truncation radii unfixed, though bound to scaling relations. We use the formalism from \citet{brimioulle2013} which states $\sigma \propto L^{0.25}$; this gives $R_{E} \propto L^{0.5}$ and $r_{200} \propto L^{0.40}$, where $L$ is found using F814W magnitudes. For our range models, we also include scatter in these relations (see Sec. \ref{sec:range}).

The large-scale halos are modeled as softened isothermal ellipsoids, or PIEMD. They are intended to account for mass in the cluster that is not found in galaxies, i.e. dark matter or ICM. This density profile is common among the parametric modeling teams since it fits cluster cores well, though we note that the magnifications it produces are not as reliable at large radii and thus should not be trusted past the strong-lensing region. We determine the number of halos to put into the model by finding the largest number of halos for which the fit improvement is statistically relevant. For the fairly relaxed cluster Abell S1063, we found that the data was well fit with only two large-scale halos, while the more complicated systems of MACS J0717 and Abell 370 required four; the other three clusters have three halos in their mass models. We also include external shear to account for any asymmetries in mass outside of the modeled region.

\subsubsection{Line-of-sight galaxies}

In cluster lens modeling, it is common to ignore effects due to LOS galaxies. One reason is that it is computationally more expensive to model with multiple lensing planes as opposed to a single plane since the lens equation becomes recursive. Another reason is that, since the LOS galaxies are much less massive than the cluster, they contribute far less to the lensing potential. However, there are cases in which large, bright foreground galaxies are very close to lensed images, or otherwise seem as if they would be important to include. One example is a foreground galaxy in MACS J0416: a source behind the cluster is lensed into a long thin arc by the cluster and shows further deflection as the arc passes through the foreground galaxy, indicating that the galaxy's mass is indeed affecting the light bending. Another example of an important LOS galaxy is the large foreground galaxy near the core of MACS J0717, which is brighter than the brightest cluster galaxy (BCG).

A modeling team may wish to include galaxies such as these, but still only have a single lensing plane for computational purposes. To accomplish this, the mass of the foreground or background galaxy is scaled such that the deflection in the single-plane model approximates what the true deflection in a multi-plane model would be (see Eq. \ref{eq:sp-approx}). Since the galaxy will only have a small effect on the lensing potential, akin to a perturber, this approximation is generally regarded to be ``close enough''. Since our team has the machinery to include multiple lens planes, we seek to quantify the effect this approximation has on cluster lens models. 

We choose which LOS galaxies to include in our models based on the brightness of the galaxies and proximity to known images. These galaxies are also treated as spherical pseudo-Jaffe models with priors placed on their Einstein radii using the same scaling relations as the cluster members. We note that LOS galaxies may have varying morphologies which will cause them to have different normalizations to the scaling relation. We determine which normalization from \citet{brimioulle2013} to use based on whether the galaxy is primarily red or blue; also based on that work (and references therein), we evolve the galaxies for the single-plane models as $L \approx (1+z)$.

\subsubsection{Range models}\label{sec:range}
In addition to the fiducial models for each cluster, we also create range models which sample the uncertainties. These are produced by perturbing the image positions by 0.5'' and re-optimizing, starting from the fiducial model. We choose this value since it is comparable to the RMS produced by our models. 

This is not the typical method of sampling the uncertainties, which is usually done with the Markov Chain Monte Carlo technique. However, it is useful in that it easily allows us to include multiple sources of uncertainty. In addition to perturbing the the image positions, we also account for uncertainties in the photometric redshifts for our LOS galaxies by varying the assigned redshift by sampling from a Gaussian distribution. We are also able to account for scatter in the mass-luminosity relation, a source of uncertainty that is not included by any other parametric modeling teams. We add Gaussian noise with a standard deviation of 0.1 dex in $\log_{10}R_E$ and 0.03 dex in $\log_{10}{r_{200}}$ when assigning mass to the galaxies. 

\section{Modeling Results}

\subsection{Abell 2744}

\begin{figure*}
\centering
\includegraphics[width=0.7\textwidth]{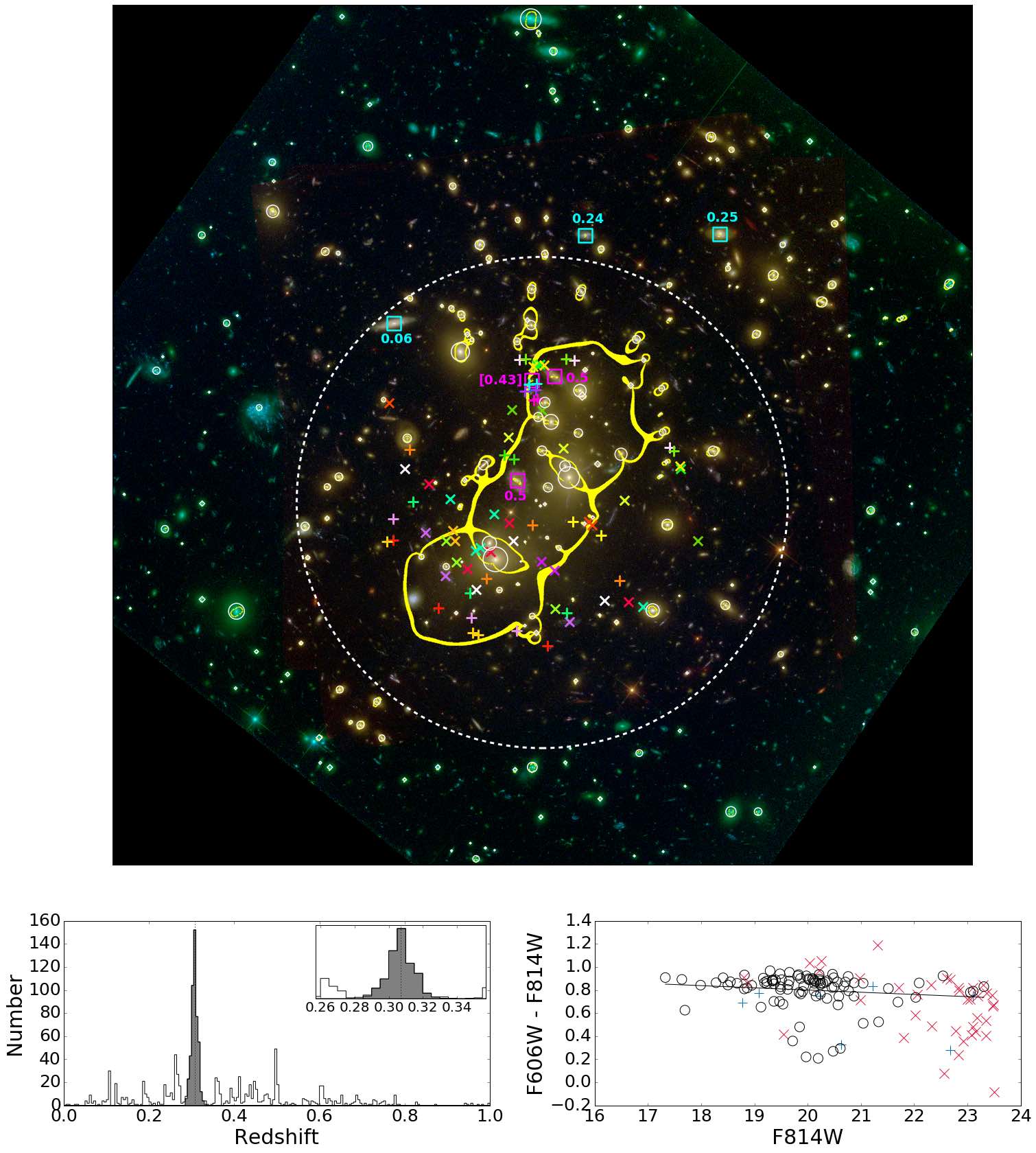}
\caption{\emph{Top:} HST color image of cluster Abell 2744 \citep[produced using Trilogy,][]{coe2012}. Cluster members are overplotted in white circles (diamonds) if they were spectroscopically (photometrically) selected, where the radius of the marker is proportional to Einstein radius. LOS foreground (background) galaxies included in the model are shown in cyan (magenta) with redshifts labelled; brackets specify photometric redshifts, otherwise they are spectroscopic. Image positions are marked with $\times$ or $+$ in varying colors; images in the same system have the same marker and color. The dashed white circle has a radius of 1 arcminute and is centered on the epicenter of the image positions. Critical curves for the 2D model at a source redshift of $z=9$ are plotted in yellow. The panel is 3.5' on a side.
\emph{Bottom:} Left shows the redshift histogram with the cluster clearly residing at a redshift $z=0.308$; the grey area shows the width of our 3-sigma cut. The color-magnitude diagram is shown on the right; black circles indicate spectroscopically-selected cluster members, while the blue $+$ (red $\times$) shows the foreground (background) galaxies.}
\label{fig:a2744skymap}
\end{figure*}

Also known as Pandora's cluster, this field was added to the Abell galaxy cluster catalog with the Southern sky update \citep{abell1989}. It is thought to have recently undergone two mergers, the primary one being line-of-sight, based on X-ray and optical studies \citep{kempner2004,owers2011,merten2011}. Evidence for this includes the large offset which exists between the peak of the X-rays and the positions of the cluster members. This conclusion is also bolstered by another study \citep{mann2012}, that found five galaxies with equal brightness (down to measurement uncertainty) in the field. These galaxies could be BCGs from the merging systems, especially since there is another peak in the X-ray data near two of these five galaxies. 

While this is clearly a complicated system, we note that we do not model all of it. Since it is quite large, some of it is outside of the HST field-of-view; we only focus on modeling that which is covered. Some modeling teams from the first round of modeling included this other region, but none of the teams in the latest round (version 4) did. The area not included is to the northwest, which is also where many modeling teams (including our own) place another halo, though ours is largely unconstrained.

This was the first field observed by HST for the HFF program in October-November of 2013 and May-July in 2014. It is located at a redshift of $z=0.308$ and has been observed previously in multiple surveys, providing publicly available data to determine cluster membership, as well as photometry needed for lens models. A recent spectroscopic survey \citep{mahler2018} was also concluded, which increased the number of lensed images with spectroscopic redshifts from less than 10 to 83. Note the medal ranking in that paper is separate from the ranking here; a system labeled gold in that paper may not have received the same ranking when all HFF modeling teams graded it. There are also some images that were added after the voting had concluded and thus did not have the amount of votes needed to obtain a \textsc{gold} ranking. 

Further, we choose not to include three systems that are classified \textsc{gold} by the HFF teams. The systems 5 and 105 consist of an arc which is clearly lensed, but is so thin that there is little discernible structure. This makes it unclear as to where to place the positions of the images or, indeed, how many images or systems there are. We also exclude system 64, since we believe 64.1 may be a merging pair of two images. In total, we use 71 images from 24 galaxies as listed in Table \ref{tab:images_a2744}; their positions in the cluster are shown in Fig. \ref{fig:a2744skymap} as + or $\times$, with consistent colors across families of images. 

The spectroscopic and photometric catalogs used to create our mass models are noted in the references of Table \ref{tab:galaxies_a2744}. From these catalogs, our cluster member selection yields 92 spectroscopically selected members, as well as an additional 163 based on photometry. The redshift histogram and color-magnitude diagram used in the selection are shown in the bottom panels of Fig. \ref{fig:a2744skymap}. The peak in the histogram clearly indicates a galaxy cluster at that redshift, but there is interestingly quite a bit of structure elsewhere. In particular, there are peaks at $z\sim0.25$ and $z\sim0.5$. We also see, as expected, a difference in the color-magnitude diagram between the different populations of cluster members, which generally lie close to the relation, and LOS galaxies, which have more spread.

Cluster members are shown in white in Fig. \ref{fig:a2744skymap}, marked as either circles or diamonds depending on if they were selected spectroscopically or photometrically; the radius of a marker is proportional to the galaxy's $R_E$. The six foreground and background galaxies included in the model are boxed in cyan and magenta, respectively, with measured redshifts labelled. The foreground galaxies were primarily included due to their size. The background galaxies, on the other hand, are close to images; they also sit much closer to the core of the cluster, and thus to the critical curves. 

These background galaxies, particularly the two spirals, have an effect on the models and drive the difference between the 2D and 3D models. The main cluster halo has a slightly larger $R_E$ in the 2D model than in the 3D one, causing higher magnifications across the cluster in the 2D maps versus the 3D maps. Interestingly, these two background galaxies are also at the same redshift, which happens to correspond to the small peak at $z\sim0.5$ seen in the redshift histogram. We do not know if these galaxies belong to a bound structure. If they do and we are missing LOS mass in our model, the model would have to shift the masses of the small- and large-scale halos to accommodate it. It may be that the cluster halo can be less massive if the galaxies can account for some of that missing mass, as we see in our 3D model. It would be interesting in future work to delve into this and determine if there are indeed line-of-sight structures we need to take into account, in addition to line-of-sight galaxies.

This cluster required three halos to model the field in addition to the galaxies. Two are positioned near the cluster core where many of the bright galaxies are seen. These main halos have mostly very similar parameters between the 2D and 3D models, with ellipticity and position angle being particularly similar. The only differences lie in the Einstein radius parameter of the first halo and the scale radii of both halos. The third halo is not near the cluster core, but instead lies to the west. Both models agree on where to place the halo with the position fairly constrained, though this is likely an artifact of modeling since there are no images in that region to actually constrain it. 

The 2D and 3D models differ on whether or not this third halo is diffuse or has a critical curve, which is determined by the ratio of the scale radius $R_S$ to Einstein radius $R_E$; if $R_E/R_S < 2$, there will not be a critical curve, but rather a diffuse area of higher magnification. While both 2D and 3D models have comparable values for the Einstein radius, the 3D model prefers a much smaller scale radius such that the halo appears diffuse while the halo in the 2D model does have a critical curve.

\begin{figure}
\centering
\includegraphics[width=0.3\textwidth]{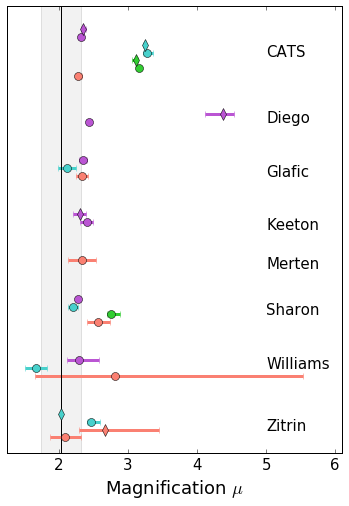}
\caption{Predicted magnifications for HFF14Tom (SN Tomas) in Abell 2744 from various teams. Colors denote which version the models are: v1 (red), v2 (green), v3 (blue), or v4 (purple); diamonds denote that they are from version ``x.1" or, in our case, 3D. The measured magnification from \citet{rodney2015} is shown at the black vertical line with errors shown in grey. We note that some models used the supernova as a constraint and thus are not a blind test.}
\label{fig:SNTom}
\end{figure}

Overall, both models fit the data well with image plane RMS values of 0.41" and 0.38" for the 2D and 3D models respectively. We choose not to use the constraint of HFF14Tom \citep[SN Tomas;][]{rodney2015}, a singly imaged type Ia supernova found to be brighter than supernovae at similar redshifts, indicating that it had been magnified by the cluster. Instead, we use it as a check to see how our model's magnification compares to the observed magnification. Our 2D model predicts a magnification of 2.41$\pm$0.10, while our 3D model predicts a slightly lower value of 2.31$\pm$0.09; the observed magnification is 2.03 $\pm$ 0.29. Our values are slightly higher than the observed, but within the error.  This is a common result amongst the teams which modeled this field, as shown in Fig. \ref{fig:SNTom}. Almost all models predict a magnification higher than what was measured, but are very close to being within the uncerainty. Note that each point includes an error bar from the submitted range models, but some bars (particularly for the parametric teams) are smaller than the size of the marker and thus not visible.

\subsection{MACS J0416.1-2403}

\begin{figure*}
\centering
\includegraphics[width=0.7\textwidth]{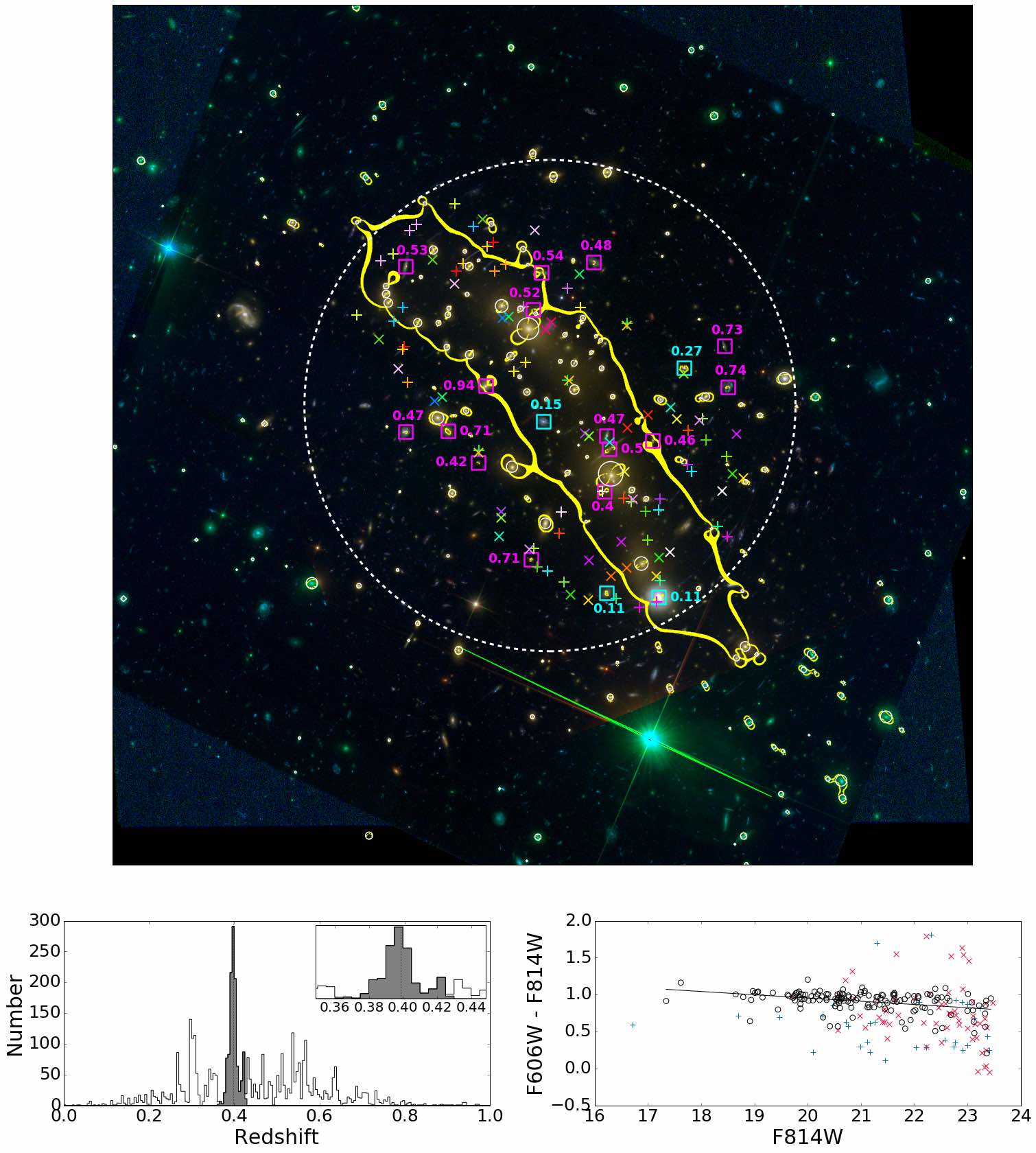}
\caption{Similar to Fig. \ref{fig:a2744skymap} for MACS J0416.}
\label{fig:m0416skymap}
\end{figure*}

A field from the Massive Cluster Survey \citep[MACS;][]{ebeling2001}, this cluster was the second observed by the HFF project in January-February and July-September of 2014. This cluster is likely undergoing a merger as evidenced by the distinctly double-peaked X-ray map and the duality of its BCGs \citep{mann2012}. It is also elongated on the sky, providing ample area for lensed images to appear. This cluster, at a redshift of $z=0.396$, has the highest number of spectroscopically confirmed images out of the six Frontier Fields.

For our modeling, we use 95 spectroscopically confirmed images from 35 sources, as listed in Table \ref{tab:images_m0416}. Most of these images got their redshifts from a survey done with VLT/MUSE by \citet{caminha2017}, adding to the work done by \citet{jauzac2014} which found redshifts for 10 systems. Though the new MUSE survey found 102 images with redshifts, we do not include all of them. Some of these images did not receive \textsc{gold} rankings, while other systems only had redshifts for one image but not the counter images. 

We also do not include two systems that were classified as \textsc{gold}. System 5 includes an image blended with a galaxy for which we did not have photometric data. System 29, which is a set of three images around a galaxy, was excluded due to the majority of its constraining power going to the Einstein radius of the specific galaxy rather than the cluster as a whole. In general, the images we did include consist of sets of triply-imaged systems across the cluster. This provides tight constraints on the position of the lateral critical curve, i.e. on the side of the cluster. The northern and southern ends, however, are less tightly constrained and more likely to be affected by, e.g., cluster members or LOS galaxies. Though this is true for many fields, it is most apparent in MACS J0416 due to its very elongated nature. 

Our cluster members are selected using publicly available catalogs as listed in Table \ref{tab:galaxies_m0416}. After applying our cuts on the data, we find the member sample shown in Fig. \ref{fig:m0416skymap}, which consists of 146 spectroscopically confirmed members with 61 additional photometrically selected members. Nineteen LOS galaxies are also included; this is a higher number than the six seen in Abell 2744, but we had a similar methodology for choosing which galaxies to include. The difference then can be attributed to a combination of the wider area over which the images in this cluster are spread, differences in data completeness, and cosmic variance. In the redshift histogram of Fig. \ref{fig:m0416skymap}, it is clear that there are more LOS galaxies in the data for this field than in the previous one. 

The LOS galaxy that most strongly affects our models is also the most obvious one: a bright, large foreground galaxy in the southern half of the cluster. As we will see in the following section, there is a clear difference in image magnifications between models that do and do not include LOS galaxies, mostly driven by this galaxy in particular. When the galaxy is included but scaled to the lens plane as in our 2D model, the magnifications around the galaxy are still quite different than in a 3D model where the galaxy is placed at its true redshift. 

In addition to galaxy-scale halos, our model consists of three large-scale halos, two near the cluster core and another to the northeast. While only the two in the core are required to fit the morphology of the lensed images, a diffuse third halo provides a significant improvement to the fit as also found by other modeling teams. This third halo required modest priors on its ellipticity, but was otherwise allowed to vary freely, as were the other halos. Our final fiducial models are able to reproduce image positions relatively well with an RMS of 0.53" for the 2D best-fit model and 0.55" for the 3D model. 

\subsection{MACS J0717.5+3745}

\begin{figure*}
\centering
\includegraphics[width=0.7\textwidth]{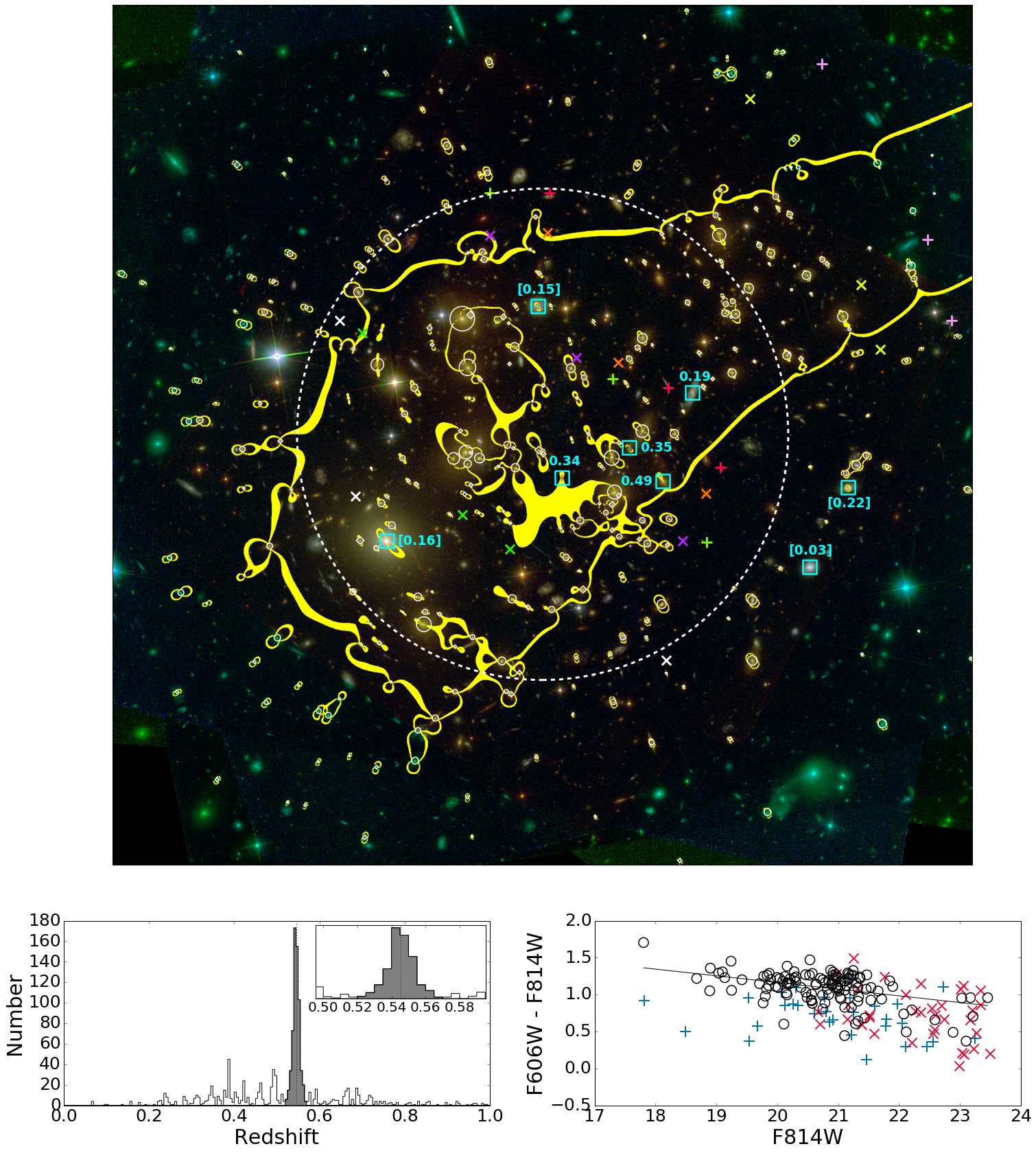}
\caption{Similar to Fig. \ref{fig:a2744skymap} for MACS J0717.}
\label{fig:m0717skymap}
\end{figure*}

This cluster, found in the MACS survey \citep{ebeling2001}, is one of the most massive clusters known at $z>0.5$. It has the highest redshift out of the sample at $z=0.545$, making it harder to carry out a spectroscopic or photometric survey to the same depth and completeness as in the other, lower redshift clusters. It is also a very complicated cluster: it was classified as the most disturbed system known at $z>0.5$ based on the large amount of structure in X-ray data in \citet{ebeling2007}. There is some evidence that it is being fed by a filament \citep{ebeling2004}, possibly causing a strange, elongated spur in its mass distribution. It was observed for the HFF in October 2013, September-November in 2014, and February-March in 2015.

Though it is clearly a very large and massive cluster, seemingly perfect for lensing background galaxies, it has the smallest amount of spectroscopically confirmed images out of the HFF survey: only 29 images from 9 sources, as listed in Table \ref{tab:images_m0717} and shown in Fig. \ref{fig:m0717skymap}. This small number does not indicate a lack of candidate or non-\textsc{gold}-ranked images however:  \citet{kawamata2016} for example use 173 images in their model. This low number of \textsc{gold} images is largely due to a lack of a dedicated spectroscopic survey as was done in the other HFF clusters.  

Our model includes 300 total cluster members, which are listed in Table \ref{tab:galaxies_m0717}. Of these, 121 are spectroscopically selected from the clear peak in the redshift histogram shown in the bottom left panel of Fig. \ref{fig:m0717skymap}. The galaxies have a broader color distribution than we see in the other HFF clusters, as shown in the bottom right panel of the figure. We thus use 3-sigma clipping on the colors of the spectroscopically selected galaxies, which we in turn use to create our color-magnitude relation. From this, our cuts yield 179 photometrically selected galaxies. 

We also include eight LOS galaxies, all of which are in the foreground. Half of these did not have spectroscopic redshifts, but we deem them important enough to include with photometric redshifts. This adds an additional source of error that we deal with by assigning a value to the galaxy from a given redshift distribution in the range models. In particular, one of these galaxies with only a photometric redshift is also the brightest galaxy in the field. It is almost the same color as the other cluster members, but has a photometric redshift of $z=0.155\pm0.03$ from CLASH \citep{postman2012,molino2017} and Subaru/Suprimecam imaging \citep{medezinski2013}. Between our 2D and 3D models, we find a large difference around this galaxy.

Due to the complexity of this cluster, our model requires at least four halos to fit the data, with three halos near the core of the cluster and another to the northwest. Based on our relatively high RMS values of 0.79" for both the 2D and 3D, we suspect that this field requires more halos to properly model its complex nature. However, we find that a fifth halo is not well constrained. Since our goal for this round of models was to only use spectroscopically confirmed images, we present our four halo models in this paper. Future work could involve including photometrically selected candidate images to bolster constraints, which would allow for increasing the complexity of our models. A more complete spectroscopic survey done on candidate images would be even better, but that of course would require dedicated observing time.

\subsection{MACS J1149.5+2223}

\begin{figure*}
\centering
\includegraphics[width=0.7\textwidth]{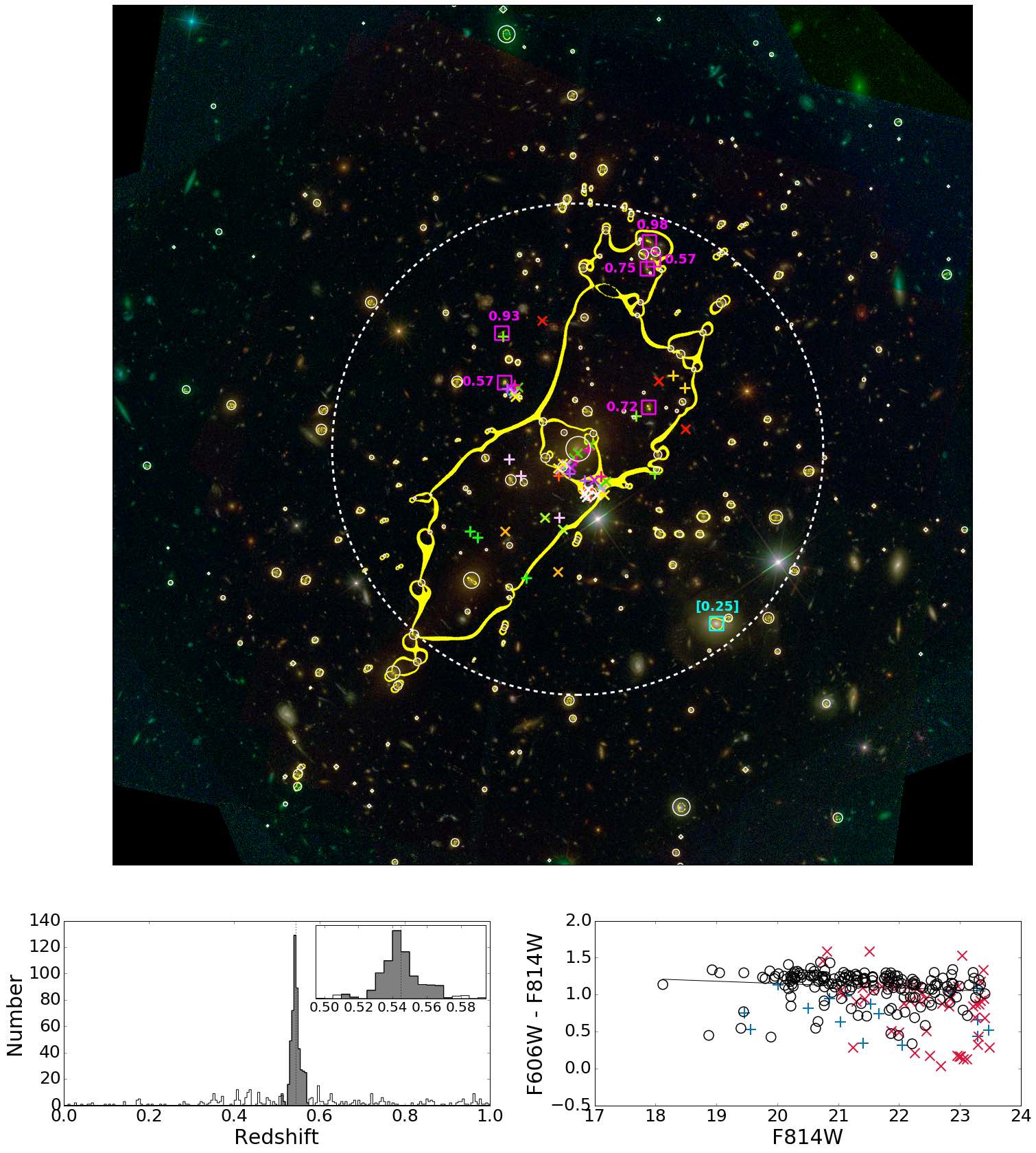}
\caption{Similar to Fig. \ref{fig:a2744skymap} for MACS J1149.}
\label{fig:m1149skymap}
\end{figure*}

This field includes another fairly high redshift cluster at $z=0.543$ and is also likely undergoing a merger, as evidenced by an elongated mass distribution. It is host to the multiply-imaged Type II SN Refsdal, which was found in a triply-lensed galaxy \citep[see][]{rodney2016, treu2016}. In addition to the three galaxy images, the supernova was in an arm of its host galaxy that happened to be further lensed into a cross formation around a cluster member galaxy. It was an exciting test of gravitational lensing because most models predicted that another image of the supernova would appear later. This particular image (SX) had a longer time delay than the other images (S1-S4) due to the lensing geometry and mass distribution of where its light travelled through the cluster. Models could then be used to predict when and where the other image would appear, creating a true blind test of cluster lensing. Most models were able to constrain the location remarkably well, to around an arcsecond, but the time delay predictions were less precise \citep{kelly2016}. This field was observed for the Frontier Fields in November 2013, November 2013-January 2014,  April 2014, and April-May 2015.

Many papers have focused on this cluster due to SN Refsdal and its triply-imaged host galaxy. However, the cluster is lacking in other spectroscopically confirmed images, with only 22 images from 9 sources ranked \textsc{gold} by the HFF modeling teams. The images we include in our model are listed in Table \ref{tab:images_m1149}. We do not include one system (9) and its four images since it is $\sim\!1.75$' away from the cluster core and is being primarily lensed by a single galaxy. In order to bolster our constraints, we add 3 extra images that were ranked \textsc{silver}: 5.2, 13.2, and 13.3. We also add constraints from seven systems of knots within the SN Refsdal host galaxy, as identified by \citet{kawamata2016}. For the SN itself, we use the positions of the five images (S1-S4 and SX), as well as a weak constraint on the time delay seen between S1 and SX, measured to be 345 $\pm$ 35 days \citep{kelly2016}. Our total number of position constraints is 53, but from only 8 sources. It is important to note that these lensed knot positions will help to constrain the model, but will only be most useful locally. 

Our cluster member selection is done using spectroscopic catalogs from a variety of sources as listed in Table \ref{tab:galaxies_m1149}. Our selection cuts yield 179 spectroscopically selected cluster members and an additional 78 based on photometry. We add seven LOS galaxies to the model, six of which are background galaxies. Three of these are found very close to images, while another three are close to a grouping of cluster members to the northwest. The single foreground galaxy included is not particularly close to images, but it is fairly large, so we include it at its photometric redshift, which we vary in the range models assuming a Gaussian distribution. 

We fit the data with a model consisting of three large-scale halos in addition to the galaxy-scale halos. Two of these are near the core of the cluster, while a third resides to the north, near a grouping of cluster members. Our 2D and 3D models have similar RMS values of 0.30" and 0.31", respectively, which are our lowest values in the HFF set. 

\subsection{Abell S1063}
\begin{figure*}
\centering
\includegraphics[width=0.7\textwidth]{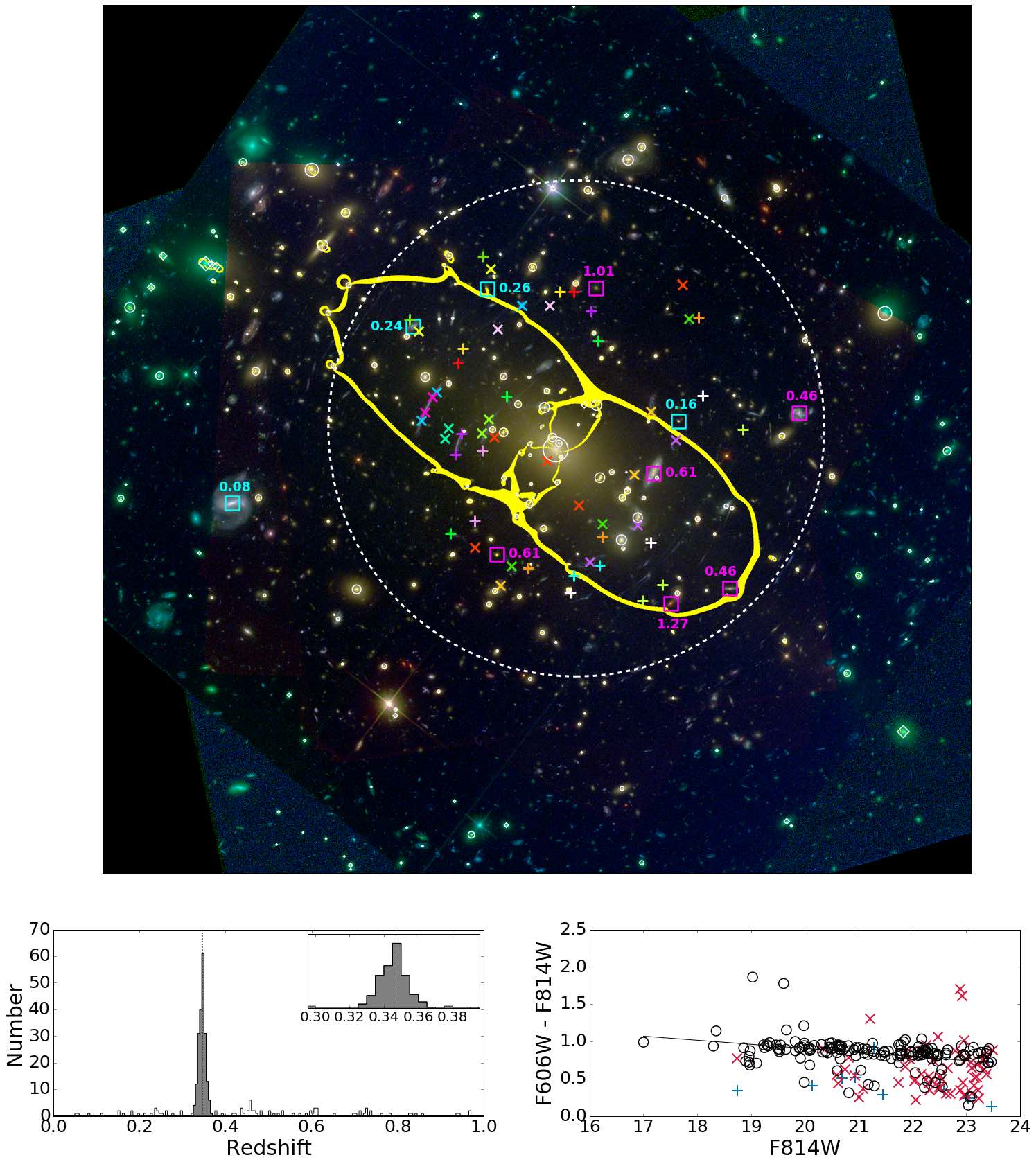}
\centering
\caption{Similar to Fig. \ref{fig:a2744skymap} for Abell S1063}
\label{fig:a1063skymap}
\end{figure*}

This cluster at $z=0.348$ seems less complicated than the others in the HFF sample. For example, the galaxy at the center is clearly the BCG, unlike other clusters which have multiple galaxies of relatively equal brightness. Yet it may be more complicated than it appears: dynamical studies have shown that it could be going through a merging event \citep{gomez2012}. Our models are able to fit the data with only two halos, the fewest of the HFF clusters. It was observed by HST in October 2014, October \& November 2015, and April \& May 2016. 

Our data, both for the cluster members and images, come from a large number of sources. The cluster member galaxies are listed in Table \ref{tab:galaxies_a1063}, while images are listed in Table \ref{tab:images_a1063}. This cluster has a relatively small number of constraints: 50 images from 19 systems. We find 153 spectroscopically confirmed members and an additional 101 using photometry. Nine LOS galaxies are added to the model, but they seem to have little effect on the models. 

Our model includes two large-scale halos, one near the BCG and another smaller halo in the southwest; we place modest priors on the ellipticity and position angle of the smaller one. We also find that the models have a difficult time properly reproducing the image near the BCG, thus, like other teams, we let that galaxy depart from the scaling relation used for cluster members. Our 2D and 3D models fit the data well and have RMS values of 0.33" and 0.34" respectively.

\subsection{Abell 370}

\begin{figure*}
\centering
\includegraphics[width=0.7\textwidth]{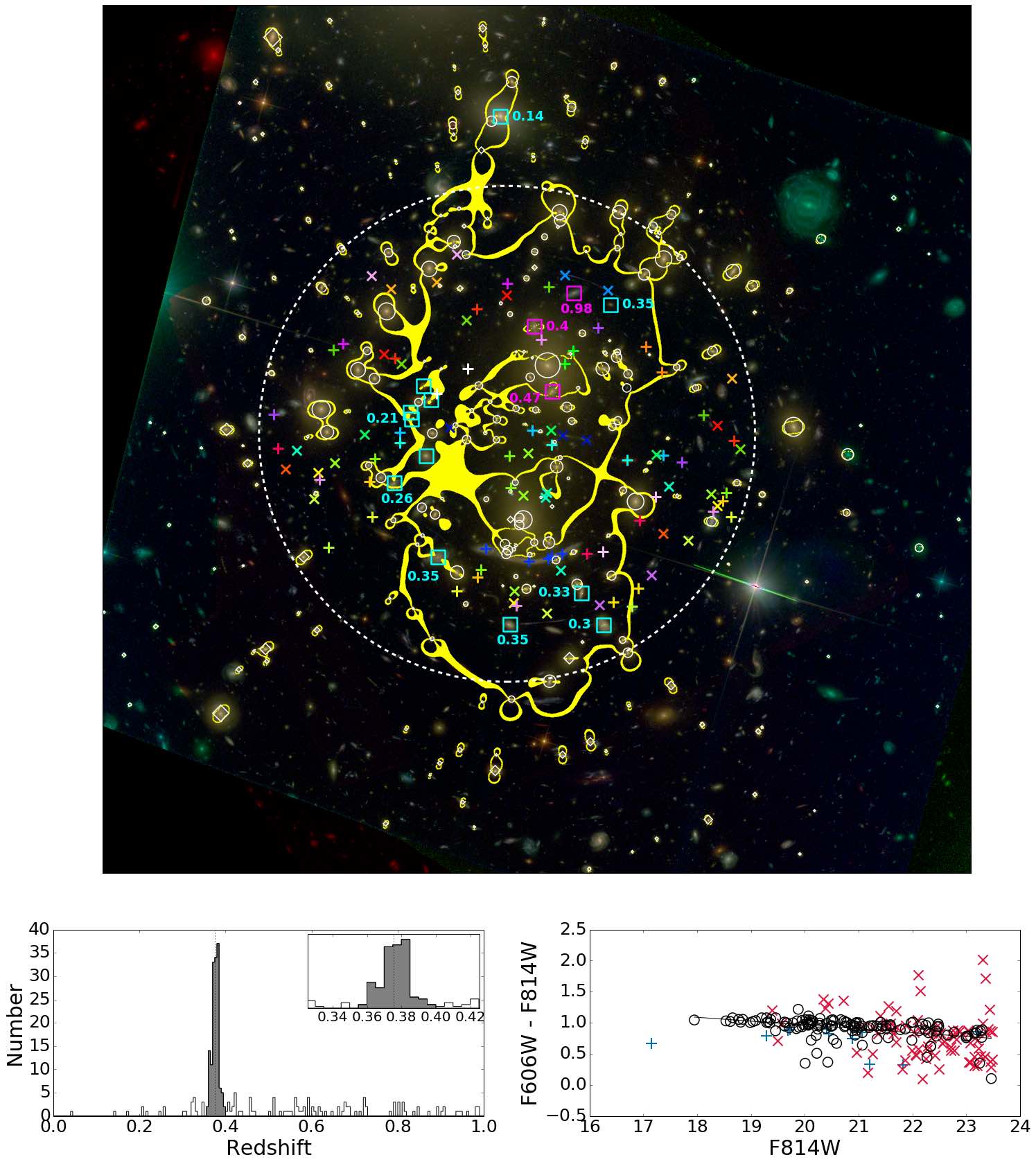}
\caption{Similar to Fig. \ref{fig:a2744skymap} for Abell 370. Due to crowding in the eastern cluster of foreground galaxies, four galaxies are not labelled. Those without a label have a redshift of $z=0.33$.}
\label{fig:a370skymap}
\end{figure*}

The sixth and final cluster in the HFF project is a well known lensing field at $z=0.375$, having been one of the first galaxy clusters in which a giant arc was observed \citep{soucail1987,lynds1989}. It was observed for the HFF program in July \& December 2015, as well as January-February and July-September 2016. 

As it was the first lensing cluster to be found, Abell 370 has been studied extensively \citep[][etc.]{defilippis2005,lah2009,holz2012}. This prior knowledge does not necessarily make the cluster easier to model, however, as it has quite a complicated mass distribution. It is clear that it is undergoing a merging event based on the two large but equally bright galaxies, indicative of two systems coming together. It also has a clump of foreground galaxies just to the east, which further complicates the critical curves and caustics. 

Since this complex system has been studied for so long, there is a rich sample of lensed images with spectroscopic redshifts. For our models we use 92 images from 30 systems, as listed in Table \ref{tab:images_a370}. We do not use two systems labelled as \textsc{gold}: 37 (using the \citealt[][]{lagattuta2017} numbering scheme) and 15 (from the \citealt[][]{diego2016} scheme). Both of these are primarily lensed by individual galaxies, and thus do not offer strong constraints on the cluster. We also treat the images from systems 7 and 10 (\citealt[][]{lagattuta2017} scheme) as if they are from a single source since they share a common redshift and the positioning of the images points to a single source.

Our model includes 129 spectroscopically selected cluster members, with an additional 127 photometrically-selected galaxies, as shown in Table \ref{tab:galaxies_a370}. This table also includes the fifteen LOS galaxies that were included in the model, all of which have spectroscopic redshifts. Many of the foreground galaxies included sit in the eastern clump. Interestingly, most of these galaxies have the same redshift; our models also place a large-scale halo in this area. We see no other obvious peaks in the histogram besides the cluster in Fig. \ref{fig:a370skymap}, but we note that it has the least populated redshift histogram out of the six HFF clusters and thus may be missing a possible foreground structure or other line-of-sight structure.

This cluster has proved complicated to model, even with the large number of constraints. We find the models require four large-scale halos in order to achieve a reasonable fit to the data. The model places two halos near the BCGs and another near the clump of foreground galaxies, while the fourth resides west of the cluster core. Our final 2D and 3D models have RMS values of 0.74" and 0.73" respectively. We increase the uncertainties on image positions in the range models for this cluster from 0.5" to 0.7" so that they are comparable to the RMS, in order to properly sample the errors. 

\section{Quantifying LOS Effects using Mock Data} \label{sec:2dvs3d}

In order to test what effects LOS galaxies may have on our models of the HFF sample, we seek to make a comparison between single-plane and multi-plane models. In order to do so, we use mock data. With this technique, we know exactly how close a model gets to reproducing the true results; this would be impossible using actual data, since the true underlying mass distribution is unknown. In this way, we can do a controlled test where we use the same techniques and fit to the same data to see how one change to the input model affects the results. Any differences in these results should be caused solely by the difference in the models, i.e. treatment of LOS galaxies. Specifically, we measure how well the models reproduce the ``true'', i.e. mock, image positions and magnifications to quantify systematic errors introduced by using single-plane approximations to fit LOS galaxies. 

The mock data start with the images positions predicted by our fiducial 3D models of the clusters. Similar to the range models described in Sec. \ref{sec:range}, we perturb these positions randomly by drawing from a 2D Gaussian distribution. In this case, however, we use the HST position uncertainty $\sigma=0.06$''. We then have 100 sets of mock image positions to which we can fit our models. We test two single-plane models which either do (2D) or do not (2D-noLOS) include LOS galaxies scaled to the lens plane. We also test a multi-plane model (3D) as a way to isolate the effects of statistical scatter. We then compare the values from these models to the ``true'' values of the mock model.

\begin{table*}
\centering
 \begin{tabular}{M{2.2cm} | M{1.5cm} M{1.8cm} M{1.5cm} | M{2.2cm} M{2.2cm} M{2.2cm}} 
 \hline
 & & Position & & & $\mu/\mu_\mathrm{{mock}}$ & \\
   \vspace{0.10cm}  & &  RMS (arcsec) &  & \multicolumn{3}{c}{median (68\% CI) } \\
   \vspace{0.10cm}Cluster & 2D-noLOS & 2D & 3D & 2D-noLOS & 2D & 3D \\ 
 \hline\hline
  \vspace{0.25cm} Abell 2744 & 0.15 & 0.13 & 0.06 & 1.06 (1.00,1.13) & 1.05 (1.01,1.09) & 1.00 (0.98,1.02) \\ 

 \vspace{0.25cm} MACS J0416 & 0.33  & 0.12 & 0.06 & 0.94 (0.86,1.05) & 1.00 (0.93,1.06) & 1.00 (0.98,1.02) \\ 

 \vspace{0.25cm} MACS J0717 & 0.14 & 0.09 & 0.07 & 0.99 (0.90,1.04) & 0.99 (0.94,1.02) & 1.00 (0.98,1.02) \\ 

 \vspace{0.25cm} MACS J1149 & 0.08 & 0.06 & 0.06 & 0.97 (0.92,1.06) & 1.02 (0.99,1.06) & 1.00 (0.97,1.03) \\

  \vspace{0.25cm} Abell S1063 & 0.08 & 0.07 & 0.06 & 1.01 (0.99,1.03)  & 1.01 (0.99,1.03) & 1.00 (0.99,1.01) \\  

  \vspace{0.25cm} Abell 370 & 0.35 & 0.11 & 0.09 & 0.99 (0.87,1.14) & 1.01 (0.99,1.06) & 1.00 (0.98,1.03) \\ 
 \hline
\end{tabular}
\caption{For each cluster, we use the mock image positions with 100 realizations of noise ($\sigma=0.06''$). Columns 2-4 give the median RMS values for the three types of models. In columns 5-7, we give the median magnification relative to the mock data, along with the 68\% confidence interval.}
\label{tab:magtable}
\end{table*}

\subsection{Effects on image positions} 

\begin{figure*}
\centering
\includegraphics[width=0.7\textwidth]{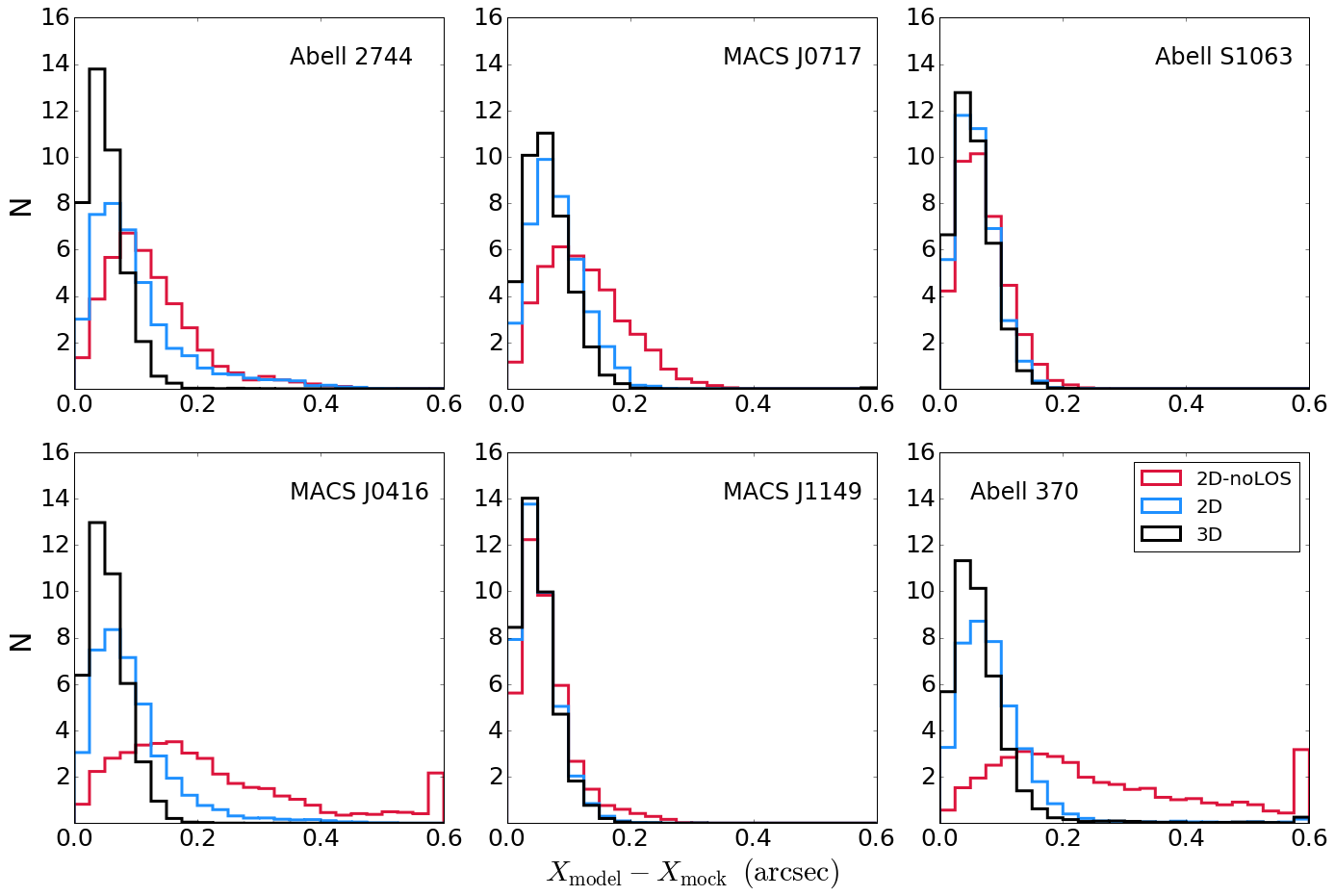}
\caption{Distances between predicted image positions and the ``true'' image positions of the mock model for the six fields. Distances greater than the edge value, 0.6, are stacked in the last bin. Width in the 3D histogram (black) is due to added positional uncertainty. Not including LOS galaxies (2D-noLOS; red) typically shifts the histogram to higher distances. Models with LOS galaxies scaled to the cluster redshift (2D; blue) typically have smaller distances. }
\centering
\label{fig:2d3ddist}
\end{figure*}

A common metric used to describe a model is how well it can reproduce the image positions of observed lensed galaxies. Most current models can produce root-mean-square (RMS) offsets of less than 1 arcsecond, and often smaller. Among our six HFF models, we have an average RMS of $\sim$0.5". Just ten years ago, these offsets were on the order of a few arcseconds; clearly there has been a great deal of improvement. However, this RMS is still much larger than the measurement uncertainty of $\sim$0.06" on image positions in HST data. 

Past studies \cite[e.g.][]{dalal2005,jullo2010,host2012} have found that mass along the LOS to a cluster can affect a lens model, including causing position offsets of up to $1-2$". It is clear that large scale structure could affect models, but individual LOS galaxies would have a smaller impact; indeed, \citet{caminha2016,chirivi2017} found that the RMS increases due to LOS galaxies could be around $0.1-0.3"$ in two fields, Abell S1063 and MACS J0416 respectively. The study of the former cluster was done with toy models, while the latter included known LOS galaxies in the field, similar to the methodology we have applied to all six fields.

In Fig. \ref{fig:2d3ddist}, we show the distance between predicted and mock image positions for each of the fields; RMS values for the three models are also shown in Table \ref{tab:magtable}. This histograms for the 3D models are driven by the statistical noise in the mock data, whereas the histograms for the 2D and 2D-noLOS models are broader because these models do not fully capture the LOS effects. All fields show some improvement in the 2D model over the 2D-noLOS, though not necessarily at the same level: MACS J0416 and Abell 370 are much improved while MACS J1149 and Abell S1063 show little change. The improvement of MACS J0416 and Abell 370 come primarily from cutting down the long tail in their distributions, which extends towards larger values.

From both the figure and table, it is clear that the results somewhat vary among the fields. Nonetheless, it is expected that adding LOS galaxies to the model, even scaled to the lens plane, would improve the fit and this is indeed what we see. Averaged across the six fields, the RMS for the 2D-noLOS and 2D models is 0.18" and 0.10", respectively, while the 3D models have a median RMS of 0.07", very close to the added uncertainty. For MACS J0416, we recover similar results to \citet{chirivi2017}, which found an RMS of 0.33'' for the "SP" (single-plane) model, which did not include LOS galaxies. While these are fairly small effects, they could still be a non-negligible portion of the error budget given that it is now well under 1".

It is important to note that small RMS values do not necessarily correlate with similar lensing products. While it is certainly a positive indication to have a model with a low RMS, one must keep in mind that it does not mean that the models are precisely recreating the true mass of the system, nor does it mean that the deliverables (e.g. image magnifications) are as accurate as a low RMS might suggest. This has been noted in other works \cite[e.g.][]{johnson2016,priewe2017} and is perhaps not surprising: the image positions themselves are the constraints used, thus the models are penalized if they cannot reconstruct those positions. This is not true for image magnifications, which are unknown in almost all cases; the only way in which a magnification would be known is for lensed supernovae, which are relatively rare. The shape of a lensed image may also tell one something indirectly about the magnification. Though many modeling teams only constrain their models with image positions, some do include pixel reconstructions \cite[e.g.][]{diego2016b}. In \cite{raney2019b}, we show that this is indeed a problem when comparing HFF models made by various teams: even though RMS values are similar, magnification maps can vary significantly.

\subsection{Effects on image magnifications}

\begin{figure*}
\centering
\includegraphics[width=0.7\textwidth]{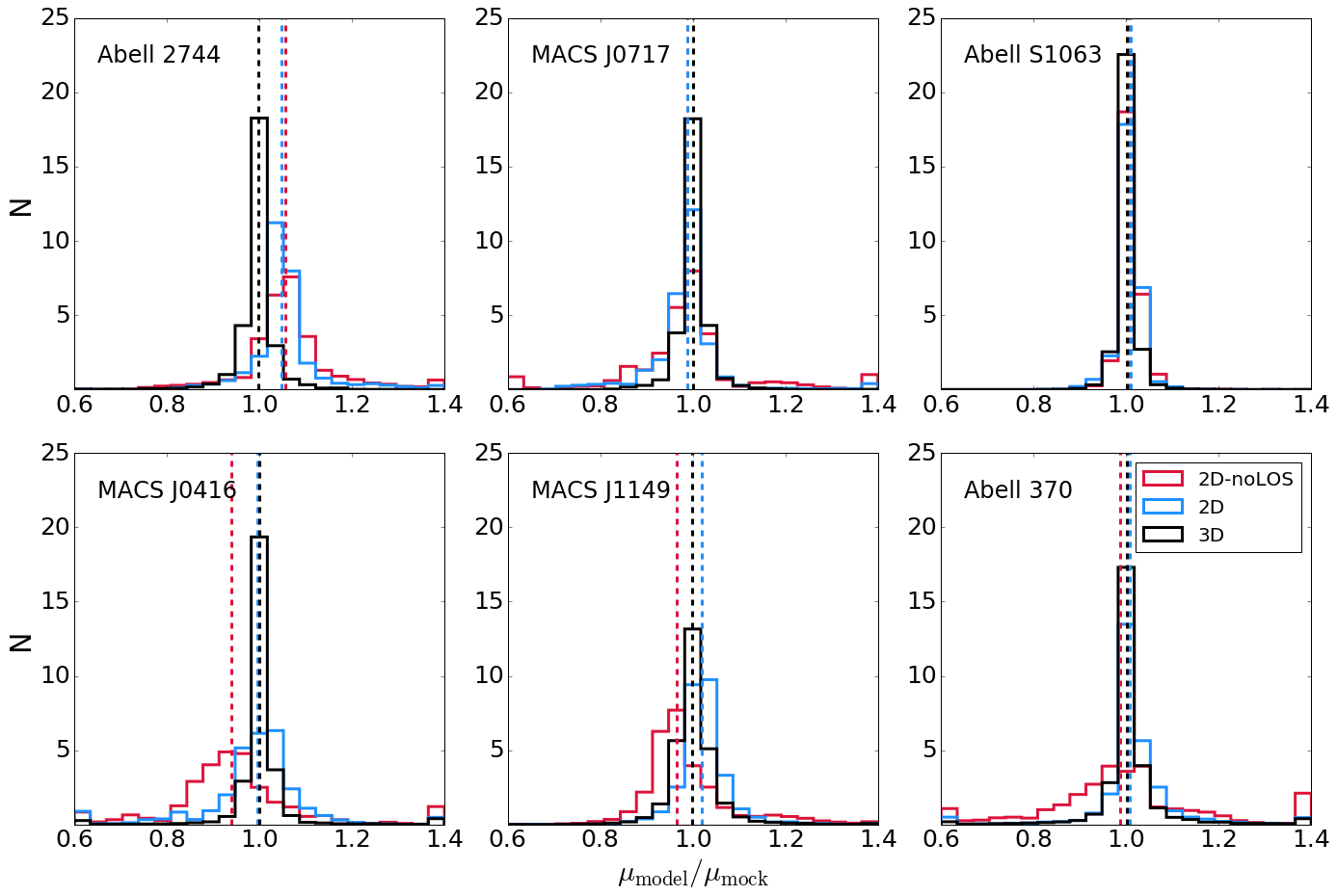}
\caption{Similar to Fig. \ref{fig:2d3ddist}, but showing magnification ratios of the predicted flux vs. the ``true'' image flux of the mock model. Median values are indicated by the dashed vertical lines. Note: values greater than those at the edges are stacked in the edge bin. Not including LOS galaxies (2D-noLOS; red) can introduce shifts or widen the distribution of magnifications; these shifts may either be high or low. Including LOS galaxies scaled to the cluster redshift (2D; blue) can lead to a correction to the bias and a tightening of the distribution in some cases, but not all. Though the 3D models all have a median ratio of 1.00, some width is present in the distribution.}
\centering
\label{fig:2d3dmag}
\end{figure*}

Magnification is an important quantity in cluster lens modeling. This is especially true for the HFF program, which sought to find distant galaxies and populate the high redshift end of the luminosity function. Work by \citet{daloisio2014} found varying large scale structure along the LOS could change magnifications by $10-30$\%. \citet{chirivi2017} found that different treatments of LOS galaxies in MACS J0416 caused biases towards lower magnifications. As with image position offsets, we expect the effects from individual LOS galaxies on magnification to be smaller than from LSS. However, any bias in magnification caused by using single-plane models could be problematic, even more so now than before the HFF observing campaign. As the number of confirmed lensed images increases and the models are better constrained, statistical errors shrink and cause systematic errors to become more prevalent. 

We see in Fig. \ref{fig:2d3dmag} and Table \ref{tab:magtable} that a shift in median magnification does exist in some of the clusters, though the direction and strength of the shift runs the gamut among the six fields. Abell 2744 and MACS J0416 are great examples of this. They are displaced in different directions: Abell 2744 towards higher magnifications in single-plane vs. multi-plane models, while MACS J0416 is biased towards lower magnification. Abell 2744 does not show a big improvement in magnification bias between the 2D-noLOS and 2D models, except for a slight tightening in the distribution in the latter case. MACS J0416, on the other hand, does show a clear improvement when LOS galaxies are added to the model. In some fields, there is no systematic shift. Abell S1063 and Abell 370, for example, both have median magnification ratios very close to unity for all models. We see, however, that the 2D model has a significantly tighter distribution than the 2D-noLOS model in Abell 370.

We note that the values reported in Table \ref{tab:magtable} are just the median bias among images spread out across the cluster; it says nothing about how large the bias may be in certain areas. For example, not including LOS galaxies at all may produce a global bias in magnification which may decrease when LOS galaxies are added to the cluster redshift. However, a local bias may still be present in the 2D model near the LOS galaxies which would not appear in the the histogram since only a small number of images would be affected. One must examine the magnification maps to determine if this is the case for a particular cluster.

It is not surprising that a bias was found in MACS J0416 since it has a large, bright foreground galaxy in the strong lensing region. Indeed, most teams account for this particular galaxy in their models by scaling it to the lens plane as we do in the 2D models. However, the results from Abell 2744 are more unexpected. The changes in magnification are caused primarily by two background spirals in the strong lensing region which lie close to the critical curves. Abell 370 is also an interesting case in that it doesn't show a bias, but the distribution of magnification ratios is significantly broader when LOS galaxies are omitted. This field has a large number of LOS galaxies in the strong lensing, similar to MACS J0416. However, it is different in that many of the galaxies are in the foreground, at the same redshift, and clumped together, which could be why the magnifications are affected differently.

\section{Conclusions and Future Work} \label{sec:results}

The Hubble Frontier Fields have provided a wealth of data for the lensing and greater astrophysics communities. In this paper, we have used these data in order to conduct the first comprehensive study of magnification biases due to line-of-sight galaxies in cluster lensing. In order to do this, we have created the first set of 3-dimensional mass models of all six Frontier Fields, in addition to two single-plane models. 

In order to determine the effects of LOS galaxies, we created a mock multi-plane model for each field and fit it with two single-plane models, either with or without LOS galaxies scaled to the lens plane, and a multi-plane model. These models were created with the same inputs, i.e. same number of halos and cluster members, and were optimized to fit the 3D data in the same way. This allowed us to quantify any bias that would be introduced from modeling a multi-plane system using only single-plane models. Our results from that can be summarized as the following:
\begin{itemize}
\item Not including LOS galaxies at all in mass models can cause offsets in image positions around 0.19" averaged across the six fields, and up to $\sim0.35$". With these galaxies included in a single-plane model through approximations, the offsets decrease to around 0.10". 
\item LOS galaxies can have an effect on magnifications, especially if massive. These effects can be important near the galaxy, i.e. within a radius of $10-20$ arcseconds, as is seen in MACS J0416. Global shifts in magnification can also occur, as evidenced by the results in Abell 2744, specifically if the galaxy (or galaxies) sits close to a critical curve.
\item Shifts in magnifications resulting from LOS galaxies can be towards either higher or lower magnifications. This can sometimes be improved by adding the LOS galaxies scaled to the lens plane, but not all the time. 
\item Models without LOS galaxies do not always show biases; it is dependent on the particulars of the field. Further, even if there is no shift, there may still be a large broadening of the magnification ratio distribution, as seen in Abell 370. 
\item A shared low RMS value between models does not necessarily mean that the products of the models, e.g. magnifications, will be similar. This is in part due to the fact that image positions are used to constrain the models, whereas magnifications are not since they are unknown in almost all cases. 
\end{itemize}

The most obvious takeaway is that the results vary widely among the clusters, likely due (at least in part) to cosmic variance. Abell 2744 shows a $6\%$ shift towards higher magnifications when the LOS galaxies are not included, but MACS J0416 shows a similar shift in the opposite direction. Abell 370 does not have a bias, but does have a much wider spread of magnification ratios in models without LOS galaxies. The magnification ratios in MACS J0416 are much improved when LOS galaxies are included, but ratios in Abell 2744 are improved only marginally.  

The last bullet point regarding RMS values is not a new result, but one that is important to keep in mind when comparing results. The idea that two models can produce similar RMS values with different magnifications may seem similar to the classic mass sheet degeneracy (MSD) where different models can predict the same exact image positions and yet have different magnifications. This is not precisely what is at work here since that particular degeneracy should be broken in cluster lensing due to multiple source redshifts. However, clearly some similar effect is at play here, though one that works at a more local level. Perhaps it is due to the large number of components in any given cluster lens model. Most degeneracy studies have been for simple lenses, which is certainly not the case for most of the HFF sample. This problem will need to be addressed in the future if cluster lensing is to be used in more careful studies. 

A caveat to our results here is that we are only quantifying the effects of particular LOS galaxies. There is also the question of whether or not there could be any other line-of-sight effects that we are not covering in our models, e.g. due to actual structures along the line of sight. For example, \citet{williams2017} found evidence for LOS structure in MACS J0717. We found in Abell 2744 that two galaxies at the same redshift (which also corresponded to a peak in the redshift histogram) affected our models on a global scale rather than the local scale seen in the other fields. In addition, our model of Abell 370 requires a fourth halo in the vicinity of a clump of foreground galaxies at the same redshift. Could this be due to the fact that these galaxies, both in Abell 2744 and Abell 370, are just a small part of a larger structure? This is an interesting question, but one that we leave for future work.

\section*{Acknowledgements}

We acknowledge support from Hubble Frontier Field Lensing Support through contract STScI-49745 from the Space Telescope Science Institute, which is operated by NASA under Contract No. NAS5-26555. All version models were obtained via the Mikulski Archive for Space Telescopes (MAST) and the web-based lens model tool. We thank Dan Coe and Keren Sharon for organizing the useful discussions among all modeling teams and contributors.  

%%%%%%%%%%%%%%%%%%%%%%%%%%%%%%%%%%%%%%%%%%%%%%%%%%



\bibliographystyle{mnras}
\bibliography{bib} % if your bibtex file is called example.bib

%%%%%%%%%%%%%%%%%%%%%%%%%%%%%%%%%%%%%%%%%%%%%%%%%%

%%%%%%%%%%%%%%%%% APPENDICES %%%%%%%%%%%%%%%%%%%%%

\appendix

\section{Galaxy catalogs}\label{sec:galcat}
\begin{table*}
\caption{Galaxies in the field Abell 2744. References are as follows:
1 = \citet{owers2011}.
2 = GLASS \citep{schmidt2014,treu2015}.
3 = ASTRODEEP \citep{castellano2016,merlin2016}.
4 = \citet{coe2015}.
5 = Subaru/Suprimecam imaging \citep{okabe2008, okabe2010a, okabe2010b}.
Note: a flag of $-1$ in the $z$ column denotes that it is a cluster member which was photometrically selected.
The full table is available in machine-readable format in the online journal.
A portion is shown here for guidance regarding its form and content.
}
\label{tab:galaxies_a2744}
\begin{tabular}{cccccll}
	\hline
	RA ($^\circ$) & Dec ($^\circ$)  & F606W & F814W & $z$ & Status & References \\
	\hline
	$  3.556336$ & $-30.387017$ & $20.43\pm0.00$ & $19.62\pm0.00$ & $ 0.311$ & member-spec    & 1,4,5      \\
	$  3.559037$ & $-30.410663$ & $20.91\pm0.00$ & $20.12\pm0.00$ & $ 0.298$ & member-spec    & 1,4,5      \\
	$\vdots$ &&&&&&\\
	$  3.596758$ & $-30.400513$ & $23.05\pm0.01$ & $22.19\pm0.00$ & $-1$ & member-phot    & 2,3,4,5    \\
	$  3.588820$ & $-30.410721$ & $21.56\pm0.00$ & $20.65\pm0.00$ & $-1$ & member-phot    & 2,3,4,5    \\
	$\vdots$ &&&&&&\\
	$  3.574393$ & $-30.383654$ & $19.86\pm0.00$ & $19.09\pm0.00$ & $ 0.255$ & los-foreground & 1,2,3,4,5  \\
	$  3.580700$ & $-30.371392$ & $20.98\pm0.00$ & $20.23\pm0.00$ & $ 0.239$ & los-foreground & 1,4,5      \\
	$\vdots$ &&&&&&\\
	$  3.552814$ & $-30.399780$ & $21.07\pm0.00$ & $20.04\pm0.00$ & $ 0.688$ & los-background & 1,4,5      \\
	$  3.563749$ & $-30.384118$ & $22.19\pm0.00$ & $21.80\pm0.00$ & $ 0.356$ & los-background & 1,4        \\
	$\vdots$ &&&&&&\\
	\hline
\end{tabular}
\end{table*}

\begin{table*}
\caption{Galaxies in the field MACS J0416. References are as follows:
1 = \citet{balestra2016}.
2 = \citet{ebeling2014}.
3 = GLASS \citep{schmidt2014,treu2015}.
4 = ASTRODEEP \citep{castellano2016, merlin2016}.
5 = CLASH \citep{postman2012,molino2017}.
6 = \citet{coe2015}.
7 = Subaru/Suprimecam imaging \citep{umetsu2011}.
Note: a flag of $-1$ in the $z$ column denotes that it is a cluster member which was photometrically selected.
The full table is available in machine-readable format in the online journal.
A portion is shown here for guidance regarding its form and content.
}
\label{tab:galaxies_m0416}
\begin{tabular}{cccccll}
	\hline
	RA ($^\circ$) & Dec ($^\circ$)  & F606W & F814W & $z$ & Status & References \\
	\hline
	$ 64.032873$ & $-24.106531$ & $20.99\pm0.00$ & $20.42\pm0.00$ & $ 0.391$ & member-spec    & 1,7            \\
	$ 64.038611$ & $-24.107521$ & $23.42\pm0.01$ & $22.69\pm0.01$ & $ 0.394$ & member-spec    & 1,7            \\
	$\vdots$ &&&&&&\\
	$ 64.037071$ & $-24.090017$ & $23.31\pm0.01$ & $22.51\pm0.00$ & $-1$ & member-phot    & 4,5,7          \\
	$ 64.021156$ & $-24.089567$ & $22.11\pm0.00$ & $21.26\pm0.00$ & $-1$ & member-phot    & 4,5,7          \\
	$\vdots$ &&&&&&\\
	$ 64.023464$ & $-24.100845$ & $21.78\pm0.01$ & $21.17\pm0.00$ & $ 0.353$ & los-foreground & 1,5,7          \\
	$ 64.045665$ & $-24.105219$ & $21.89\pm0.00$ & $21.26\pm0.00$ & $ 0.303$ & los-foreground & 1,7            \\
	$\vdots$ &&&&&&\\
	$ 64.039029$ & $-24.107523$ & $23.42\pm0.02$ & $22.68\pm0.01$ & $ 0.844$ & los-background & 1,7            \\
	$ 64.044334$ & $-24.095366$ & $22.56\pm0.01$ & $21.63\pm0.00$ & $ 0.570$ & los-background & 1,4,5,7        \\
	$\vdots$ &&&&&&\\
	\hline
\end{tabular}
\end{table*}

\begin{table*}
\caption{Galaxies in the field MACS J0717. References are as follows:
1 = \citet{ebeling2014}.
2 = GLASS \citep{schmidt2014,treu2015}.
3 = CLASH \citep{postman2012,molino2017}.
4 = \citet{coe2015}.
5 = SDSS DR13 \citet{york2000,albareti2017}.
6 = Subaru/Suprimecam imaging \citep{medezinski2013}.
Note: a flag of $-1$ in the $z$ column denotes that it is a cluster member which was photometrically selected.
The full table is available in machine-readable format in the online journal.
A portion is shown here for guidance regarding its form and content.
}
\label{tab:galaxies_m0717}
\begin{tabular}{cccccll}
	\hline
	RA ($^\circ$) & Dec ($^\circ$)  & F606W & F814W & $z$ & Status & References \\
	\hline
	$109.353789$ & $ 37.735569$ & $20.66\pm0.01$ & $19.77\pm0.00$ & $ 0.536$ & member-spec    & 1,3,4,6      \\
	$109.355199$ & $ 37.743205$ & $22.23\pm0.01$ & $20.96\pm0.00$ & $ 0.547$ & member-spec    & 1,3,4,6      \\
	$\vdots$ &&&&&&\\
	$109.394237$ & $ 37.779913$ & $23.96\pm0.04$ & $23.03\pm0.01$ & $-1$ & member-phot    & 3,4,6        \\
	$109.399562$ & $ 37.774545$ & $24.34\pm0.04$ & $23.33\pm0.01$ & $-1$ & member-phot    & 3,4          \\
	$\vdots$ &&&&&&\\
	$109.353758$ & $ 37.737308$ & $18.99\pm0.00$ & $18.49\pm0.00$ & $ 0.069$ & los-foreground & 1,3,4,6      \\
	$109.355201$ & $ 37.747004$ & $21.35\pm0.01$ & $20.61\pm0.00$ & $ 0.284$ & los-foreground & 1,3,4,6      \\
	$\vdots$ &&&&&&\\
	$109.363160$ & $ 37.733868$ & $21.98\pm0.01$ & $21.39\pm0.00$ & $ 0.575$ & los-background & 1,3,4,6      \\
	$109.378340$ & $ 37.758656$ & $22.25\pm0.02$ & $20.97\pm0.00$ & $ 0.576$ & los-background & 1,3,4,6      \\
	$\vdots$ &&&&&&\\
	\hline
\end{tabular}
\end{table*}

\begin{table*}
\caption{Galaxies in the field MACS J1149. References are as follows:
1 = \citet{ebeling2014}.
2 = GLASS \citep{schmidt2014,treu2015}.
3 = SDSS DR13 (spectroscopy).
4 = SDSS DR13 (photometry) \citep{york2000,albareti2017}.
5 = CLASH \citep{postman2012,molino2017}.
6 = \citet{coe2015}.
7 = Subaru/Suprimecam imaging (Based on data collected at Subaru Telescope by PI K. Umetsu (in prep) and archival imaging obtained from Subaru-Mitaka-Okayama-Kiso Archive (SMOKA), which is operated by the Astronomy Data Center, National Astronomical Observatory of Japan.).
Note: a flag of $-1$ in the $z$ column denotes that it is a cluster member which was photometrically selected.
The full table is available in machine-readable format in the online journal.
A portion is shown here for guidance regarding its form and content.
}
\label{tab:galaxies_m1149}
\begin{tabular}{cccccll}
	\hline
	RA ($^\circ$) & Dec ($^\circ$)  & F606W & F814W & $z$ & Status & References \\
	\hline
	$177.370263$ & $ 22.393006$ & $22.50\pm0.00$ & $21.30\pm0.00$ & $ 0.546$ & member-spec    & 1,5,6,7        \\
	$177.371565$ & $ 22.410314$ & $21.48\pm0.01$ & $20.22\pm0.00$ & $ 0.553$ & member-spec    & 1,4,5,6,7      \\
	$\vdots$ &&&&&&\\
	$177.376918$ & $ 22.394621$ & $23.34\pm0.02$ & $22.13\pm0.01$ & $-1$ & member-phot    & 4,5,6,7        \\
	$177.404133$ & $ 22.428316$ & $21.71\pm0.01$ & $20.72\pm0.00$ & $-1$ & member-phot    & 4,5,6,7        \\
	$\vdots$ &&&&&&\\
	$177.375157$ & $ 22.386274$ & $21.80\pm0.01$ & $20.86\pm0.00$ & $ 0.463$ & los-foreground & 1,4,5,6,7      \\
	$177.382634$ & $ 22.427029$ & $21.15\pm0.10$ & $20.02\pm0.00$ & $ 0.514$ & los-foreground & 1,4,5,7        \\
	$\vdots$ &&&&&&\\
	$177.394687$ & $ 22.378622$ & $23.41\pm0.02$ & $22.31\pm0.01$ & $ 0.680$ & los-background & 1,4,5,6,7      \\
	$177.411484$ & $ 22.429689$ & $21.53\pm0.09$ & $21.24\pm0.20$ & $ 1.227$ & los-background & 1,4,7          \\
	$\vdots$ &&&&&&\\
	\hline
\end{tabular}
\end{table*}

\begin{table*}
\caption{Galaxies in the field Abell S1063. References are as follows:
1 = \citet{karman2017}.
2 = \citet{karman2015}.
3 = \citet{gomez2012}.
4 = \citet{caminha2016}.
5 = GLASS \citep{schmidt2014,treu2015,grillo2015}.
6 = CLASH \citep{postman2012,molino2017}.
7 = \citet{coe2015}.
Note: a flag of $-1$ in the $z$ column denotes that it is a cluster member which was photometrically selected.
The full table is available in machine-readable format in the online journal.
A portion is shown here for guidance regarding its form and content.
}
\label{tab:galaxies_a1063}
\begin{tabular}{cccccll}
	\hline
	RA ($^\circ$) & Dec ($^\circ$)  & F606W & F814W & $z$ & Status & References \\
	\hline
	$342.167155$ & $-44.534685$ & $21.13\pm0.00$ & $20.82\pm0.00$ & $ 0.326$ & member-spec    & 1,2,6,7    \\
	$342.179379$ & $-44.527908$ & $19.76\pm0.01$ & $18.96\pm0.00$ & $ 0.328$ & member-spec    & 1,6,7      \\
	$\vdots$ &&&&&&\\
	$342.174817$ & $-44.500392$ & $20.71\pm0.01$ & $19.77\pm0.00$ & $-1$ & member-phot    & 6          \\
	$342.183579$ & $-44.499862$ & $23.28\pm0.02$ & $22.48\pm0.01$ & $-1$ & member-phot    & 6          \\
	$\vdots$ &&&&&&\\
	$342.180723$ & $-44.546572$ & $24.14\pm0.04$ & $23.41\pm0.02$ & $ 0.153$ & los-foreground & 1,2,6,7    \\
	$342.171663$ & $-44.528959$ & $23.60\pm0.02$ & $23.47\pm0.01$ & $ 0.160$ & los-foreground & 1,2,6,7    \\
	$\vdots$ &&&&&&\\
	$342.205805$ & $-44.523279$ & $22.67\pm0.01$ & $22.32\pm0.01$ & $ 0.439$ & los-background & 1,5,6,7    \\
	$342.166800$ & $-44.540196$ & $22.72\pm0.02$ & $22.15\pm0.01$ & $ 0.458$ & los-background & 1,2,5,6,7  \\
	$\vdots$ &&&&&&\\
	\hline
\end{tabular}
\end{table*}

\begin{table*}
\caption{Galaxies in the field Abell 370 References are as follows:
1 = \citet{lagattuta2017}.
2 = GLASS (spectroscopy).
3 = GLASS (photometry) \citep{schmidt2014,treu2015}.
4 = \citet{coe2015}.
5 = SDSS DR13 (spectroscopy).
6 = SDSS DR13 (photometry) \citep{york2000,albareti2017}.
Note: a flag of $-1$ in the $z$ column denotes that it is a cluster member which was photometrically selected.
The full table is available in machine-readable format in the online journal.
A portion is shown here for guidance regarding its form and content.
}
\label{tab:galaxies_a370}
\begin{tabular}{cccccll}
	\hline
	RA ($^\circ$) & Dec ($^\circ$)  & F606W & F814W & $z$ & Status & References \\
	\hline
	$ 39.975198$ & $ -1.587928$ & $22.31\pm0.00$ & $21.55\pm0.00$ & $ 0.358$ & member-spec    & 1,2,3      \\
	$ 39.978460$ & $ -1.583929$ & $22.96\pm0.00$ & $22.34\pm0.00$ & $ 0.361$ & member-spec    & 1,2,3      \\
	$\vdots$ &&&&&&\\
	$ 39.968471$ & $ -1.553844$ & $22.99\pm0.00$ & $22.19\pm0.00$ & $-1$ & member-phot    & 3,4        \\
	$ 39.974150$ & $ -1.557419$ & $21.66\pm0.07$ & $20.83\pm0.02$ & $-1$ & member-phot    & 3          \\
	$\vdots$ &&&&&&\\
	$ 39.978832$ & $ -1.575483$ & $22.15\pm0.00$ & $21.83\pm0.00$ & $ 0.207$ & los-foreground & 1,2,3,4    \\
	$ 39.980006$ & $ -1.579790$ & $21.91\pm0.00$ & $21.06\pm0.00$ & $ 0.256$ & los-foreground & 1,2,3,4    \\
	$\vdots$ &&&&&&\\
	$ 39.973514$ & $ -1.580083$ & $23.89\pm0.02$ & $23.48\pm0.01$ & $ 0.410$ & los-background & 1,3        \\
	$ 39.969400$ & $ -1.573644$ & $20.59\pm0.00$ & $19.39\pm0.00$ & $ 0.466$ & los-background & 1,2,3,4,6  \\
	$\vdots$ &&&&&&\\
	\hline
\end{tabular}
\end{table*}

\section{Image Catalogs}\label{sec:imcat}

\begin{table}
\caption{Lensed Images for Abell 2744. References are for the spectroscopic redshifts and are as follows:
1 = \citet{johnson2014},
2 = \citet{richard2014},
3 = \citet{wang2015},
4 = \citet{mahler2018}. Notes are as follows: 
a = All teams ranked the image as secure, but redshift was tentative.
b = This image did not have enough votes to obtain a ranking, but \citet{mahler2018} report a redshift.
c = This image did not have enough votes to obtain a ranking.
Since some images do not have spectroscopic redshifts that we could find, they may have no reference listed. 
In these cases, we assume the image has the same redshift as the system.  
We use \citet{jauzac2014} as the coordinate reference for all systems less than 61 and \citet{mahler2018} for those above. The full table is available in machine-readable format in the online journal.}
\label{tab:images_a2744}
\begin{tabular}{lccccc} 
	\hline
	ID & RA ($^\circ$) & Dec ($^\circ$)  & $z$ & Rank & References \\
	\hline
	1.1 & $3.597542$ & $-30.403917$ & $1.688$ & \textsc{Gold} & 4 \\
	1.2 & $3.595958$ & $-30.406822$ & $1.688$ & \textsc{Gold}  & 4 \\
	1.3 & $3.586208$ & $-30.409986$ & $1.688$ & \textsc{Gold}  & 3,4 \\
	3.1 & $3.589375$ & $-30.393875$ & $3.980$ & \textsc{Gold}  & 1,4 \\
	3.2 & $3.588792$ & $-30.393803$ & $3.980$ & \textsc{Gold}  & 1,4 \\
	4.1 & $3.592125$ & $-30.402633$ & $3.572$ & \textsc{Gold}  & 4 \\
	4.2 & $3.595625$ & $-30.401622$ & $3.572$ & \textsc{Gold}  & 4 \\
	4.3 & $3.580417$ & $-30.408925$ & $3.572$ & \textsc{Gold}  & 2,4 \\
	4.4 & $3.593208$ & $-30.404914$ & $3.572$ & \textsc{Gold}  & 4 \\
	4.5 & $3.593583$ & $-30.405106$ & $3.572$ & \textsc{Gold}  & 2,4 \\
	6.1 & $3.598542$ & $-30.401800$ & $2.016$ & \textsc{Gold}  & 2,3,4 \\
	6.2 & $3.594042$ & $-30.408011$ & $2.016$ & \textsc{Gold}  & 3,4 \\
	6.3 & $3.586417$ & $-30.409372$ & $2.016$ & \textsc{Gold}  & 2,3,4 \\
	8.1 & $3.589708$ & $-30.394339$ & $3.975$ & \textsc{Gold}  & 4 \\
	8.2 & $3.588833$ & $-30.394222$ & $3.975$ & \textsc{Gold}  & 4 \\
	10.1 & $3.588417$ & $-30.405878$ & $2.655$ & \textsc{Gold}  & 4 \\
	10.2 & $3.587375$ & $-30.406481$ & $2.655$ & \textsc{Gold}  & 4 \\
	18.1 & $3.590750$ & $-30.395561$ & $5.660$ & \textsc{Gold}  & 4 \\
	18.2 & $3.588375$ & $-30.395636$ & $5.660$ & \textsc{Gold}  & 4 \\
	18.3 & $3.576125$ & $-30.404475$ & $5.660$ & \textsc{Gold}  & 3,4 \\
	22.1 & $3.587917$ & $-30.411611$ & $5.283$ & \textsc{Gold}  & 4 \\
	22.2 & $3.600083$ & $-30.404417$ & $5.283$ & \textsc{Gold}  & 4 \\
	22.3 & $3.596542$ & $-30.409031$ & $5.283$ & \textsc{Gold}  & 4 \\
	24.1 & $3.595917$ & $-30.404483$ & $1.043$ & \textsc{Gold}  & 4 \\
	24.2 & $3.595125$ & $-30.405933$ & $1.043$ & \textsc{Gold}  & 4 \\
	24.3 & $3.587333$ & $-30.409103$ & $\hdots$ & \textsc{Silver}$^a$ & 4 \\
	26.1 & $3.593958$ & $-30.409686$ & $3.052$ & \textsc{Gold}  & 4 \\
	26.2 & $3.590375$ & $-30.410586$ & $3.052$ & \textsc{Gold}  & 4 \\
	26.3 & $3.600083$ & $-30.402969$ & $3.052$ & \textsc{Gold}  & 4 \\
	30.1 & $3.591000$ & $-30.397444$ & $1.025$ & \textsc{Gold}  & 4 \\
	30.2 & $3.586708$ & $-30.398186$ & $1.025$ & \textsc{Gold}  & 4 \\
	30.3 & $3.581917$ & $-30.401703$ & $1.025$ & \textsc{Gold}  & 4 \\
	31.1 & $3.585917$ & $-30.403167$ & $4.758$ & \textsc{Gold}  & 4 \\
	31.2 & $3.583708$ & $-30.404117$ & $4.758$ & \textsc{Gold}  & 4 \\
	33.1 & $3.584708$ & $-30.403147$ & $5.723$ & \textsc{Gold}  & 4 \\
	33.2 & $3.584417$ & $-30.403389$ & $5.723$ & \textsc{Gold}  & 4 \\
	33.3 & $3.600417$ & $-30.395111$ & $5.723$ & \textsc{None}$^b$ & 4 \\
	34.1 & $3.593417$ & $-30.410842$ & $3.785$ & \textsc{Gold}  & 4 \\
	34.2 & $3.593833$ & $-30.410725$ & $3.785$ & \textsc{Gold}  & 4 \\
	34.3 & $3.600583$ & $-30.404533$ & $3.785$ & \textsc{Gold}  & 4 \\
	37.1 & $3.589042$ & $-30.394914$ & $2.649$ & \textsc{Gold}  & 4 \\
	37.2 & $3.588758$ & $-30.394836$ & $2.649$ & \textsc{Gold}  & 4 \\
	39.1 & $3.588792$ & $-30.392531$ & $4.015$ & \textsc{Gold}  & 4 \\
	39.2 & $3.588542$ & $-30.392508$ & $4.015$ & \textsc{Gold}  & 4 \\
	39.3 & $3.577458$ & $-30.399564$ & $4.015$ & \textsc{None}$^b$ & 4 \\
	40.1 & $3.589083$ & $-30.392664$ & $4.015$ & \textsc{Gold}  & 4 \\
	40.2 & $3.588208$ & $-30.392553$ & $4.015$ & \textsc{Gold}  & 4 \\
	40.3 & $3.577542$ & $-30.399372$ & $\hdots$ & \textsc{None}$^c$\\
	41.1 & $3.599167$ & $-30.399583$ & $4.910$ & \textsc{Gold}$^a$ & 4 \\
	41.2 & $3.593558$ & $-30.407769$ & $4.910$ & \textsc{Gold}$^a$ & 4 \\
	41.3 & $3.583458$ & $-30.408500$ & $4.910$ & \textsc{Gold}$^a$ & 4 \\
	41.4 & $3.590617$ & $-30.404458$ & $4.910$ & \textsc{None}$^{b}$ & 4 \\
\end{tabular}
\end{table}

\begin{table}
\contcaption{Lensed Images for Abell 2744.}
\label{tab:images_a2744_cont}
\begin{tabular}{lccccc}
	\hline
	ID & RA ($^\circ$) & Dec ($^\circ$)  & $z$ & Rank & References \\
	\hline
	42.1 & $3.597292$ & $-30.400608$ & $3.690$ & \textsc{Gold}  & 4  \\
	42.2 & $3.590958$ & $-30.403261$ & $3.690$ & \textsc{Gold}  & 4  \\
	42.3 & $3.581583$ & $-30.408633$ & $3.690$ & \textsc{Gold}  & 4  \\
	42.4 & $3.594250$ & $-30.406389$ & $3.690$ & \textsc{Gold}  & 4  \\
	42.5 & $3.592412$ & $-30.405194$ & $3.690$ & \textsc{Gold}  & 4  \\
	61.1 & $3.595375$ & $-30.403783$ & $2.952$ & \textsc{Gold}  & 4 \\
	61.2 & $3.595250$ & $-30.404450$ & $2.952$ & \textsc{Gold}  & 4 \\
	62.1 & $3.591326$ & $-30.398643$ & $4.192$ & \textsc{Gold}  & 4 \\
	62.2 & $3.590582$ & $-30.398918$ & $4.192$ & \textsc{Gold}  & 4 \\
	63.1 & $3.582261$ & $-30.407166$ & $5.660$ & \textsc{Gold}  & 4 \\
	63.2 & $3.592758$ & $-30.407022$ & $5.660$ & \textsc{Gold}  & 4 \\
	63.3 & $3.589133$ & $-30.403419$ & $5.660$ & \textsc{Gold}  & 4 \\
	63.4 & $3.598805$ & $-30.398279$ & $5.660$ & \textsc{Gold}  & 4 \\
	47.1 & $3.590162$ & $-30.392181$ & $4.022$ & \textsc{None}$^b$ & 4 \\
	47.2 & $3.585842$ & $-30.392244$ & $4.022$ & \textsc{None}$^b$ & 4 \\
	47.3 & $3.578329$ & $-30.398133$ & $4.022$ & \textsc{None}$^b$ & 4 \\
	147.1 & $3.589679$ & $-30.392136$ & $4.022$ & \textsc{None}$^b$ & 4 \\
	147.2 & $3.586454$ & $-30.392128$ & $4.022$ & \textsc{None}$^b$ & 4 \\
	147.3 & $3.578008$ & $-30.398392$ & $4.022$ & \textsc{None}$^b$ & 4 \\
	\hline
\end{tabular}
\end{table}

\newpage

\begin{table*}
\caption{Lensed Images for MACS J0416. References are for the spectroscopic redshifts and are as follows:
1 = \citet{christensen2012},
2 = \citet{grillo2015},
3 = GLASS \citep{schmidt2014,treu2015},
4 = \citet{caminha2016}. Notes are as follows:
a = This system was ranked \textsc{Gold}, but we could find no data in the literature for a spectroscopic redshift. 
Since some images do not have spectroscopic redshifts that we could find, they may have no reference listed. 
In these cases, we assume the image has the same redshift as the system. We use \citet{jauzac2014} as the coordinate references on IDs below 28; IDs 31 and above use \citet{caminha2016} as a reference. In system 30, D15 refers to \citet{diego2015} which also served as the coordinate reference.
The full table is available in machine-readable format in the online journal.}
\label{tab:images_m0416}
\begin{tabular}{lcccccc}
	\hline
	ID & Old ID & RA ($^\circ$) & Dec ($^\circ$)  & $z$ & Rank & References \\
	\hline
1.1 & 1.1 & $64.040750$ & $-24.061592$ & $1.896$ & \textsc{Gold}  & 1 \\
1.2 & 1.2 & $64.043479$ & $-24.063542$ & $\hdots$ & \textsc{Gold}$^a$  & \\
1.3 & 1.3 & $64.047354$ & $-24.068669$ & $\hdots$ & \textsc{Gold}$^a$  & \\
2.1 & 2.1 & $64.041183$ & $-24.061881$ & $1.895$ & \textsc{Gold}  & 2,3,4 \\
2.2 & 2.2 & $64.043004$ & $-24.063036$ & $1.895$ & \textsc{Gold}  & 3,4 \\
2.3 & 2.3 & $64.047475$ & $-24.068850$ & $1.895$ & \textsc{Gold}  & 3,4 \\
3.1 & 3.1 & $64.030783$ & $-24.067117$ & $1.989$ & \textsc{Gold}  & 2,3,4 \\
3.2 & 3.2 & $64.035254$ & $-24.070981$ & $1.989$ & \textsc{Gold}  & 3,4 \\
3.3 & 3.3 & $64.041817$ & $-24.075711$ & $1.989$ & \textsc{Gold}  & 2,3,4 \\
4.1 & 4.1 & $64.030825$ & $-24.067225$ & $1.990$ & \textsc{Gold}  & 2,3 \\
4.2 & 4.2 & $64.035154$ & $-24.070981$ & $1.990$ & \textsc{Gold}  & 3 \\
4.3 & 4.3 & $64.041879$ & $-24.075856$ & $1.990$ & \textsc{Gold}  & 2,3 \\
6.1 & 7.1 & $64.039800$ & $-24.063092$ & $2.088$ & \textsc{Gold}  & 3,4 \\
6.2 & 7.2 & $64.040633$ & $-24.063561$ & $2.088$ & \textsc{Gold}  & 2,3,4 \\
6.3 & 7.3 & $64.047117$ & $-24.071108$ & $\hdots$ & \textsc{Gold}$^a$  & \\
7.1 & 10.1 & $64.026017$ & $-24.077156$ & $2.298$ & \textsc{Gold}  & 2,3 \\
7.2 & 10.2 & $64.028471$ & $-24.079756$ & $2.298$ & \textsc{Gold}  & 2 \\
7.3 & 10.3 & $64.036692$ & $-24.083901$ & $\hdots$ & \textsc{Gold}$^a$  & \\
8.1 & 11.1 & $64.039208$ & $-24.070367$ & $1.005$ & \textsc{Gold}  & 4 \\
8.2 & 11.2 & $64.038317$ & $-24.069753$ & $1.005$ & \textsc{Gold}  & 4 \\
8.3 & 11.3 & $64.034259$ & $-24.066018$ & $1.005$ & \textsc{Gold}  & 4 \\
9.1 & 13.1 & $64.027579$ & $-24.072786$ & $3.217$ & \textsc{Gold}  & 2,4 \\
9.2 & 13.2 & $64.032129$ & $-24.075169$ & $3.217$ & \textsc{Gold}  & 4 \\
9.3 & 13.3 & $64.040338$ & $-24.081544$ & $\hdots$ & \textsc{Gold}$^a$  & \\
10.1 & 14.1 & $64.026233$ & $-24.074339$ & $1.633$ & \textsc{Gold}  & 2,3,4 \\
10.2 & 14.2 & $64.031042$ & $-24.078961$ & $1.633$ & \textsc{Gold}  & 2,3,4 \\
10.3 & 14.3 & $64.035825$ & $-24.081328$ & $1.633$ & \textsc{Gold}  & 3,4 \\
11.1 & 16.1 & $64.024058$ & $-24.080894$ & $1.966$ & \textsc{Gold}  & 3 \\
11.2 & 16.2 & $64.028329$ & $-24.084542$ & $1.966$ & \textsc{Gold}  & 3 \\
11.3 & 16.3 & $64.031596$ & $-24.085769$ & $1.966$ & \textsc{Gold}  & 2,3 \\
12.1 & 17.1 & $64.029875$ & $-24.086364$ & $\hdots$ & \textsc{Gold}$^a$  &  \\
12.2 & 17.2 & $64.028608$ & $-24.085986$ & $\hdots$ & \textsc{Gold}$^a$ & \\
12.3 & 17.3 & $64.023329$ & $-24.081581$ & $2.218$ & \textsc{Gold}  & 2,3 \\
13.1 & 23.1 & $64.044546$ & $-24.072100$ & $2.094$ & \textsc{Gold}  & 3 \\
13.2 & 23.2 & $64.039604$ & $-24.066631$ & $\hdots$ & \textsc{Gold}$^a$  & \\
13.3 & 23.3 & $64.034342$ & $-24.063742$ & $2.091$ & \textsc{Gold}  & 3 \\
14.1 & 26.1 & $64.046470$ & $-24.060393$ & $3.236$ & \textsc{Gold}  & 4 \\
14.2 & 26.2 & $64.046963$ & $-24.060793$ & $3.236$ & \textsc{Gold}  & 4 \\
14.3 & 26.3 & $64.049089$ & $-24.062876$ & $3.236$ & \textsc{Gold}  & 4 \\
15.1 & 27.1 & $64.048159$ & $-24.066959$ & $\hdots$ & \textsc{Gold}$^a$  & \\
15.2 & 27.2 & $64.047465$ & $-24.066026$ & $2.107$ & \textsc{Gold}  & 3 \\
15.3 & 27.3 & $64.042226$ & $-24.060543$ & $\hdots$ & \textsc{Gold}$^a$  & \\
16.1 & 28.1 & $64.036457$ & $-24.067026$ & $0.940$ & \textsc{Gold}  & 3,4 \\
16.2 & 28.2 & $64.036870$ & $-24.067498$ & $0.940$ & \textsc{Gold}  & 3,4 \\
17.1 & 33.1 & $64.028427$ & $-24.082995$ & $5.365$ & \textsc{Gold}  & 4 \\
17.2 & 33.2 & $64.035052$ & $-24.085486$ & $5.365$ & \textsc{Gold}  & 4 \\
17.3 & 33.3 & $64.022980$ & $-24.077275$ & $5.366$ & \textsc{Gold}  & 4 \\
18.1 & 34.1 & $64.029254$ & $-24.073289$ & $5.106$ & \textsc{Gold}  & 4 \\
18.2 & 34.2 & $64.030798$ & $-24.074180$ & $5.106$ & \textsc{Gold}  & 4 \\
\end{tabular}
\end{table*}

\begin{table*}
\contcaption{Lensed Images for MACS J0416.}
\label{tab:images_m0416_cont}
\centering
\begin{tabular}{lcccccc}
	\hline
	ID & Old ID & RA ($^\circ$) & Dec ($^\circ$)  & $z$ & Rank & References \\
	\hline
19.1 & 35.1 & $64.037492$ & $-24.083636$ & $3.491$ & \textsc{Gold}  & 4 \\
19.2 & 35.2 & $64.029418$ & $-24.079861$ & $3.491$ & \textsc{Gold}  & 4 \\
19.3 & 35.3 & $64.024937$ & $-24.075016$ & $3.491$ & \textsc{Gold}  & 4 \\
20.1 & 38.1 & $64.033625$ & $-24.083178$ & $3.441$ & \textsc{Gold}  & 4 \\
20.2 & 38.2 & $64.031255$ & $-24.081905$ & $3.441$ & \textsc{Gold}  & 4 \\
20.3 & 38.3 & $64.022701$ & $-24.074589$ & $3.441$ & \textsc{Gold}  & 4 \\
21.1 & 44.1 & $64.045259$ & $-24.062757$ & $3.288$ & \textsc{Gold}  & 4 \\
21.2 & 44.2 & $64.041543$ & $-24.059997$ & $3.288$ & \textsc{Gold}  & 4 \\
21.3 & 44.3 & $64.049237$ & $-24.068168$ & $3.288$ & \textsc{Gold}  & 4 \\
22.1 & 47.1 & $64.026328$ & $-24.076694$ & $3.253$ & \textsc{Gold}  & 4 \\
22.2 & 47.2 & $64.028329$ & $-24.078999$ & $3.253$ & \textsc{Gold}  & 4 \\
23.1 & 48.1 & $64.035489$ & $-24.084668$ & $4.122$ & \textsc{Gold}  & 4 \\
23.2 & 48.2 & $64.029244$ & $-24.081802$ & $4.122$ & \textsc{Gold}  & 4 \\
23.3 & 48.3 & $64.023416$ & $-24.076122$ & $4.122$ & \textsc{Gold}  & 4 \\
24.1 & 49.1 & $64.033944$ & $-24.074569$ & $3.871$ & \textsc{Gold}  & 4 \\
24.2 & 49.2 & $64.040175$ & $-24.079864$ & $3.871$ & \textsc{Gold}  & 4 \\
25.1 & 51.1 & $64.040160$ & $-24.080290$ & $4.103$ & \textsc{Gold}  & 4 \\
25.2 & 51.2 & $64.033663$ & $-24.074752$ & $4.103$ & \textsc{Gold}  & 4 \\
25.3 & 51.3 & $64.026620$ & $-24.070494$ & $4.103$ & \textsc{Gold}  & 4 \\
26.1 & 55.1 & $64.035233$ & $-24.064726$ & $3.292$ & \textsc{Gold}  & 4 \\
26.2 & 55.2 & $64.038514$ & $-24.065965$ & $3.292$ & \textsc{Gold}  & 4 \\
27.1 & 58.1 & $64.025187$ & $-24.073582$ & $3.077$ & \textsc{Gold}  & 4 \\
27.2 & 58.2 & $64.037730$ & $-24.082390$ & $3.077$ & \textsc{Gold}  & 4 \\
27.3 & 58.3 & $64.030481$ & $-24.079220$ & $3.077$ & \textsc{Gold}  & 4 \\
28.1 & 67.1 & $64.038075$ & $-24.082404$ & $3.110$ & \textsc{Gold}  & 4 \\
28.2 & 67.2 & $64.025451$ & $-24.073651$ & $3.110$ & \textsc{Gold}  & 4 \\
28.3 & 67.3 & $64.030363$ & $-24.079019$ & $3.110$ & \textsc{Gold}  & 4 \\
30.1 & 32.1 (D15) & $64.045119$ & $-24.072336$ & $3.288$ & \textsc{Gold}  & 4 \\
30.2 & 32.2 (D15) & $64.040081$ & $-24.066730$ & $3.288$ & \textsc{Gold}  & 4 \\
31.1 & 2a & $64.050865$ & $-24.066538$ & $6.145$ & \textsc{Gold}  & 4 \\
31.2 & 2b & $64.048179$ & $-24.062406$ & $6.145$ & \textsc{Gold}  & 4 \\
31.3 & 2c & $64.043572$ & $-24.059004$ & $6.145$ & \textsc{Gold}  & 4 \\
32.1 & 6a & $64.047808$ & $-24.070164$ & $3.607$ & \textsc{Gold}  & 4 \\
32.2 & 6b & $64.043657$ & $-24.064401$ & $3.607$ & \textsc{Gold}  & 4 \\
32.3 & 6c & $64.037676$ & $-24.060756$ & $3.607$ & \textsc{Gold}  & 4 \\
33.1 & 22a & $64.030997$ & $-24.077173$ & $3.923$ & \textsc{Gold}  & 4 \\
33.2 & 22b & $64.027127$ & $-24.073572$ & $3.923$ & \textsc{Gold}  & 4 \\
34.1 & 23a & $64.035668$ & $-24.079920$ & $2.542$ & \textsc{Gold}  & 4 \\
34.2 & 23b & $64.032638$ & $-24.078508$ & $2.542$ & \textsc{Gold}  & 4 \\
35.1 & 33a & $64.032017$ & $-24.084230$ & $5.973$ & \textsc{Gold}  & 4 \\
35.2 & 33b & $64.030821$ & $-24.083697$ & $5.973$ & \textsc{Gold}  & 4 \\
36.1 & 34a & $64.027632$ & $-24.082609$ & $3.923$ & \textsc{Gold}  & 4 \\
36.2 & 34b & $64.023731$ & $-24.078477$ & $3.923$ & \textsc{Gold}  & 4 \\
37.1 & 35a & $64.033681$ & $-24.085855$ & $5.639$ & \textsc{Gold}  & 4 \\
37.2 & 35b & $64.028654$ & $-24.084240$ & $5.639$ & \textsc{Gold}  & 4 \\
37.3 & 35c & $64.022187$ & $-24.077559$ & $5.639$ & \textsc{Gold}  & 4 \\
	\hline
\end{tabular}
\end{table*}

\begin{table}
\caption{Lensed Images for MACS J0717. References are for the spectroscopic redshifts and are as follows:
1 = \citet{zitrin2009b},
2 = \citet{limousin2012},
3 = GLASS \citep{schmidt2014,treu2015},
4 = \citet{vanzella2014},
5 = \citet{richard2014},
6 = \citet{medezinski2013}. Notes are as follows:
a = This system was ranked \textsc{gold}, but we could find no data in the literature for a spectroscopic redshift. 
Since some images do not have spectroscopic redshifts that we could find, they may have no reference listed. 
In these cases, we assume the image has the same redshift as the system. 
The full table is available in machine-readable format in the online journal.}
\label{tab:images_m0717}
\begin{tabular}{lccccc}
	\hline
	ID & RA ($^\circ$) & Dec ($^\circ$)  & $z$ & Rank & References \\
	\hline
	1.1 & $109.395338$ & $37.741175$ & $\hdots$ & \textsc{Gold}$^a$ & \\
 	1.2 & $109.393826$ & $37.740092$ & $2.963$ & \textsc{Gold} & 1 \\
 	1.3 & $109.390988$ & $37.738286$ & $2.963$ & \textsc{Gold} & 1,2 \\
 	1.4 & $109.384352$ & $37.736947$ & $\hdots$ & \textsc{Gold}$^a$ & \\
 	1.5 & $109.405784$ & $37.761374$ & $\hdots$ & \textsc{Gold}$^a$  & \\
	3.1 & $109.398546$ & $37.741503$ & $1.855$ & \textsc{Gold} & 1,3 \\
	3.2 & $109.394459$ & $37.739172$ & $1.855$ & \textsc{Gold} & 1,3 \\
	3.3 & $109.407155$ & $37.753830$ & $1.855$ & \textsc{Gold} & 1,2,3 \\
	4.1 & $109.380870$ & $37.750119$ & $1.855$ & \textsc{Gold} & 1,3 \\
	4.2 & $109.376438$ & $37.744689$ & $1.855$ & \textsc{Gold} & 1,3 \\
	4.3 & $109.391094$ & $37.763300$ & $1.855$ & \textsc{Gold} & 1,3 \\
	6.1 & $109.364357$ & $37.757097$ & $2.393$ & \textsc{Gold} & 1,3 \\
	6.2 & $109.362705$ & $37.752681$ & $\hdots$ & \textsc{Gold}$^a$ & \\
	6.3 & $109.373863$ & $37.769703$ & $\hdots$ & \textsc{Gold}$^a$ & \\
	12.1 & $109.385165$ & $37.751836$ & $1.710$ & \textsc{Gold} & 1,3 \\
	12.2 & $109.377617$ & $37.742914$ & $1.710$ & \textsc{Gold} & 1,3 \\
	12.3 & $109.391219$ & $37.760630$ & $1.710$ & \textsc{Gold} & 1,3 \\
	13.1 & $109.385674$ & $37.750722$ & $2.547$ & \textsc{Gold} & 1,3 \\
	13.2 & $109.377564$ & $37.739614$ & $\hdots$ & \textsc{Gold}$^a$ & \\
	13.3 & $109.396212$ & $37.763333$ & $2.547$ & \textsc{Gold} & 1,3 \\
	14.1 & $109.388791$ & $37.752164$ & $1.855$ & \textsc{Gold} & 2,3 \\
	14.2 & $109.379664$ & $37.739703$ & $1.855$ & \textsc{Gold} & 2,3 \\
	14.3 & $109.396192$ & $37.760419$ & $1.855$ & \textsc{Gold} & 2,3 \\
	15.1 & $109.367663$ & $37.772058$ & $2.405$ & \textsc{Gold} & 2 \\
	15.2 & $109.358624$ & $37.760133$ & $\hdots$ & \textsc{Gold}$^a$ & \\
	15.3 & $109.356540$ & $37.754641$ & $\hdots$ & \textsc{Gold}$^a$ & \\
	19.1 & $109.409067$ & $37.754682$ & $6.387$ & \textsc{Gold} & 4 \\
	19.2 & $109.407728$ & $37.742741$ & $6.387$ & \textsc{Gold} & 4 \\
	19.3 & $109.381057$ & $37.731611$ & $\hdots$ & \textsc{Gold}$^a$ & \\
	\hline
\end{tabular}
\end{table}
 
\begin{table}
\caption{Lensed Images for MACS J1149. References are for the spectroscopic redshifts and are as follows:
1 = \citet{smith2009},
2 = GLASS \citep{treu2015,grillo2015,wang2017},
3 = \citet{jauzac2016},
4 = \citet{johnson2014}. Notes are as follows:
a = This image was not ranked \textsc{Gold} due to an error stating no spectroscopic redshift exists. 
Since some images do not have spectroscopic redshifts that we could find, they may have no reference listed. 
In these cases, we assume the image has the same redshift as the system. 
Coordinate references are from \citet{johnson2014} for systems 14 and under. Knot identification is from \citet{kawamata2016}. 
The full table is available in machine-readable format in the online journal.}
\label{tab:images_m1149}
\begin{tabular}{lccccc}
	\hline
	ID & RA ($^\circ$) & Dec ($^\circ$)  & $z$ & Rank & References \\
	\hline
	1.1 & $177.397000$ & $22.396000$ & $1.488$ & \textsc{Gold} & 1,2 \\
	1.2 & $177.399420$ & $22.397439$ & $1.488$ & \textsc{Gold} & 1,2 \\
	1.3 & $177.403420$ & $22.402439$ & $1.488$ & \textsc{Gold} & 1,2 \\
	2.1 & $177.402420$ & $22.389750$ & $1.892$ & \textsc{Gold} & 1,2 \\
	2.2 & $177.406040$ & $22.392478$ & $1.894$ & \textsc{Gold} & 1,2 \\
	2.3 & $177.406580$ & $22.392886$ & $1.894$ & \textsc{Gold} & 1,2 \\
	3.1 & $177.390750$ & $22.399847$ & $3.129$ & \textsc{Gold} & 1,2,3 \\
	3.2 & $177.392710$ & $22.403081$ & $3.129$ & \textsc{Gold} & 1,2 \\
	3.3 & $177.401290$ & $22.407189$ & $3.131$ & \textsc{Gold} & 1,2 \\
	4.1 & $177.393000$ & $22.396825$ & $2.949$ & \textsc{Gold} & 1,2,3 \\
	4.2 & $177.394380$ & $22.400736$ & $2.949$ & \textsc{Gold} & 1,2 \\
	4.3 & $177.404170$ & $22.406128$ & $2.949$ & \textsc{Gold} & 1,2 \\
	5.1 & $177.399750$ & $22.393061$ & $2.800$ & \textsc{Gold} & 2,3 \\
	5.2 & $177.401080$ & $22.393825$ & $\hdots$ & \textsc{Silver} & \\
	13.1 & $177.403710$ & $22.397786$ & $1.239$ & \textsc{Gold} & 2,4 \\
	13.2 & $177.402830$ & $22.396656$ & $1.252$ & \textsc{Silver}$^a$ & 2,4 \\
	13.3 & $177.400040$ & $22.393858$ & $1.233$ & \textsc{Bronze}$^a$ & 2,4 \\
	14.1 & $177.391670$ & $22.403489$ & $3.703$ & \textsc{Gold} & 2,4 \\
	14.2 & $177.390830$ & $22.402647$ & $3.703$ & \textsc{Bronze}$^a$ & 2,4 \\
	110.1 & $177.400140$ & $22.390162$ & $3.214$ & \textsc{Gold} & 2 \\
	110.2 & $177.404020$ & $22.392894$ & $3.214$ & \textsc{Gold} & 2 \\
SN1 & $177.398230$ & $22.395631$ & $1.488$ \\
SN2 & $177.397720$ & $22.395783$ & $1.488$ \\
SN3 & $177.397370$ & $22.395539$ & $1.488$ \\
SN4 & $177.397810$ & $22.395189$ & $1.488$ \\
SX & $177.400083$ & $22.396694$ & $1.488$ \\
1.2.1 & $177.396615$ & $22.396308$ & $1.488$ \\
1.2.2 & $177.398978$ & $22.397892$ & $1.488$ \\
1.2.3 & $177.403041$ & $22.402689$ & $1.488$ \\
1.2.4 & $177.397765$ & $22.398780$ & $1.488$ \\
1.2.6 & $177.398674$ & $22.398225$ & $1.488$ \\
1.13.1 & $177.396974$ & $22.396636$ & $1.488$ \\
1.13.2 & $177.398832$ & $22.397717$ & $1.488$ \\
1.13.3 & $177.403311$ & $22.402814$ & $1.488$ \\
1.13.4 & $177.397907$ & $22.398433$ & $1.488$ \\
1.16.1 & $177.397446$ & $22.396394$ & $1.488$ \\
1.16.2 & $177.399154$ & $22.397219$ & $1.488$ \\
1.16.3 & $177.403596$ & $22.402647$ & $1.488$ \\
1.17.1 & $177.398140$ & $22.396353$ & $1.488$ \\
1.17.2 & $177.399274$ & $22.396839$ & $1.488$ \\
1.17.3 & $177.403845$ & $22.402569$ & $1.488$ \\
1.19.1 & $177.396892$ & $22.395761$ & $1.488$ \\
1.19.2 & $177.399538$ & $22.397483$ & $1.488$ \\
1.19.3 & $177.403367$ & $22.402286$ & $1.488$ \\
1.19.5 & $177.399962$ & $22.397094$ & $1.488$ \\
1.23.1 & $177.396724$ & $22.395372$ & $1.488$ \\
1.23.2 & $177.399757$ & $22.397494$ & $1.488$ \\
1.23.3 & $177.403257$ & $22.402025$ & $1.488$ \\
1.23.5 & $177.400133$ & $22.397203$ & $1.488$ \\
1.30.1 & $177.398171$ & $22.395469$ & $1.488$ \\
1.30.2 & $177.398008$ & $22.395231$ & $1.488$ \\
1.30.3 & $177.397308$ & $22.395372$ & $1.488$ \\
1.30.4 & $177.397896$ & $22.395728$ & $1.488$ \\
\hline
\end{tabular}
\end{table}

\begin{table}
\caption{Lensed Images for Abell S1063. References are for the spectroscopic redshifts and are as follows:
1 = \citet{balestra2013},
2 = \citet{boone2013},
3 = \citet{richard2014},
4 = \citet{johnson2014},
5 = \citet{karman2015},
6 = \citet{caminha2016},
7 = \citet{karman2017}. Notes are as follows:
a = This image was ranked \textsc{gold}, but we could find no data for a spectroscopic redshift.  
b = This image was ranked \textsc{gold}, though no spectroscopic redshifts exists; all modelers felt confident that it was part of the lensed system.
c = This image was not reported to have a spectroscopic redshift in \citet{karman2017}, but a new reduction of the MUSE data by B. Cl\'ement detected Ly$\alpha$ (private communication). 
$\dagger$ = This system is to be read as ``system.image.knot". For 1.1.1 and 1.2.1, we use the coordinates referenced in \citet{diego2016}, while 1.1.2 and 1.2.2 reference those in \citet{caminha2016}. 
$\ddagger$ = This image was proposed in \citet{monna2014}; here we assign it the ID 12.5.
Since some images do not have spectroscopic redshifts that we could find, they may have no reference listed. 
In these cases, we assume the image has the same redshift as the system. 
For systems 44 and under (except where otherwise specified), we use \citet{diego2016} as the coordinate reference. The systems with ID's greater than that use \citet{karman2017} for the coordinate reference.
The full table is available in machine-readable format in the online journal.}
\label{tab:images_a1063}
\begin{tabular}{lccccc}
	\hline
	ID & RA ($^\circ$) & Dec ($^\circ$)  & $z$ & Rank & Refs. \\
	\hline 
	1.1.1$^\dagger$ & $342.194450$ & $-44.527003$ & $1.228$ & \textsc{Gold}  & 1,7 \\
	1.1.2$^\dagger$ & $342.195590$ & $-44.528390$ & $1.229$ & \textsc{Gold}  & 6 \\
	1.2.1$^\dagger$ & $342.195867$ & $-44.528950$ & $1.228$ & \textsc{Gold}  & 1,3,7 \\
	1.2.2$^\dagger$ & $342.194830$ & $-44.527350$ & $1.229$ & \textsc{Gold}  & 6 \\
	1.3 & $342.186421$ & $-44.521203$ & $1.228$ & \textsc{Gold}  & 1,3,4,6,7 \\
	2.1 & $342.192708$ & $-44.531189$ & $1.259$ & \textsc{Gold}  & 1,4,6,7 \\
	2.2 & $342.192125$ & $-44.529831$ & $1.259$ & \textsc{Gold}  & 1,4,6,7 \\
	2.3 & $342.179863$ & $-44.521561$ & $1.259$ & \textsc{Gold}  & 3,4 \\
	4.1 & $342.193708$ & $-44.530161$ & $1.258$ & \textsc{Gold}  & 7 \\
	4.2 & $342.193333$ & $-44.529419$ & $1.258$ & \textsc{Gold}  & 7 \\
	5.1 & $342.179208$ & $-44.523589$ & $1.397$ & \textsc{Gold}  & 1,6,7 \\
	5.2 & $342.187833$ & $-44.527311$ & $1.397$ & \textsc{Gold}  & 1,3,4,6,7 \\
	5.3 & $342.193167$ & $-44.536531$ & $1.397$ & \textsc{Gold}  & 4,7 \\
	6.1 & $342.174250$ & $-44.528331$ & $1.428$ & \textsc{Gold}  & 3,4,5,7 \\
	6.2 & $342.175833$ & $-44.532539$ & $1.428$ & \textsc{Gold}  & 5,7 \\
	6.3 & $342.188438$ & $-44.539994$ & $1.428$ & \textsc{Gold}  & 1,4,5,6,7 \\
	7.1 & $342.169375$ & $-44.527250$ & $\hdots$ & \textsc{Gold}$^a$  &  \\
	7.2 & $342.174250$ & $-44.537111$ & $1.837$ & \textsc{Gold}  & 6,7 \\
	7.3 & $342.181833$ & $-44.540500$ & $1.837$ & \textsc{Gold}  & 6,7 \\
	10.1 & $342.190238$ & $-44.529764$ & $0.729$ & \textsc{Gold}  & 7 \\
	10.2 & $342.189550$ & $-44.528842$ & $0.729$ & \textsc{Gold}  & 7 \\
	11.1 & $342.175042$ & $-44.541031$ & $3.116$ & \textsc{Gold}  & 4,5,7 \\
	11.2 & $342.173167$ & $-44.539981$ & $3.116$ & \textsc{Gold}  & 1,5,6,7 \\
	11.3 & $342.165554$ & $-44.529531$ & $\hdots$ & \textsc{Gold}$^a$ &  \\
	12.1 & $342.189042$ & $-44.530050$ & $6.107$ & \textsc{Gold}  & 7 \\
	12.2 & $342.181042$ & $-44.534619$ & $6.107$ & \textsc{Gold}  & 1,2,3,5 \\
	12.3 & $342.190875$ & $-44.537469$ & $6.107$ & \textsc{Gold}  & 1,2,3,5 \\
	12.4 & $342.171292$ & $-44.519811$ & $6.107$& \textsc{Gold}  & 1,2,3 \\
	12.5$^\ddagger$ & $342.184080$ & $-44.531620$ & $6.107$  & \textsc{Gold}  & 7 \\
	13.1 & $342.181550$ & $-44.539375$ & $4.113$ & \textsc{Gold}  & 5,7 \\
	13.2 & $342.179138$ & $-44.538678$ & $4.113$ & \textsc{Gold}  & 5,7 \\
	14.1 & $342.178833$ & $-44.535869$ & $3.118$ & \textsc{Gold}  & 6,7 \\
	14.2 & $342.187417$ & $-44.538689$ & $3.118$ & \textsc{Gold}  & 6,7 \\
	14.3 & $342.170667$ & $-44.522089$ & $\hdots$ & \textsc{Gold}$^a$ & \\
	17.1 & $342.185833$ & $-44.538850$ & $3.606$ & \textsc{Gold}  & 7 \\
	17.2 & $342.178833$ & $-44.536731$ & $3.606$ & \textsc{Gold}  & 7 \\
	17.3 & $342.169792$ & $-44.521978$ & $\hdots$ & \textsc{Gold}$^b$  & \\
	19.1 & $342.180021$ & $-44.538431$ & $1.035$ & \textsc{Gold}  & 5 \\
	19.2 & $342.175542$ & $-44.535939$ & $1.035$ & \textsc{Gold}  & 5 \\
	19.3 & $342.171917$ & $-44.530250$ & $1.035$ & \textsc{Gold}  & 5 \\
	44.2 & $342.192442$ & $-44.525069$ & $2.976$ & \textsc{Gold}  & 7 \\
	44.3 & $342.181504$ & $-44.520258$ & $\hdots$ & \textsc{Bronze}  & \\
\end{tabular}
\end{table}

\begin{table}
\contcaption{Lensed Images for Abell S1063.}
\label{tab:images_a1063_cont}
\begin{tabular}{lccccc}
	\hline
	ID & RA ($^\circ$) & Dec ($^\circ$)  & $z$ & Rank & References \\
	\hline
	93a & $342.182830$ & $-44.520280$ & $3.169$ & \textsc{Gold}  & 7 \\
	93b & $342.191960$ & $-44.524090$ & $3.169$ & \textsc{Gold}  & 7 \\
	94a & $342.189350$ & $-44.518710$ & $3.286$ & \textsc{Gold}  & 7 \\
	94b & $342.196150$ & $-44.522910$ & $3.286$ & \textsc{Gold}  & 7 \\
	98a & $342.190150$ & $-44.530930$ & $5.051$ & \textsc{Gold}  & 7 \\
	98b & $342.190850$ & $-44.535660$ & $5.051$ & \textsc{Gold}  & 7 \\
	99a & $342.183780$ & $-44.521220$ & $5.237$ & \textsc{Gold}  & 7 \\
	99b & $342.188740$ & $-44.522760$ & $5.237$ & \textsc{Gold}  & 7 \\
	100a & $342.197010$ & $-44.522121$ & $5.894$ & \textsc{Gold}  & 7 \\
	100b & $342.190100$ & $-44.517886$ & $5.894$ & \textsc{Gold}$^c$   & 7 \\
	\hline
\end{tabular}
\end{table}

\begin{table}
\caption{Lensed Images for Abell 370. References are for the spectroscopic redshifts and are as follows:
1 = \citet{soucail1988},
2 = \citet{richard2014},
3 = GLASS \citep{schmidt2014,treu2015},
4 = \citet{lagattuta2017},
5 = \citet{lagattuta2019}. Notes are as follows:
a = This image had a lower quality (1 or 2) flag on the spectroscopic redshift. 
Since some images do not have spectroscopic redshifts that we could find, they may have no reference listed. 
In these cases, we assume the image has the same redshift as the system. 
We use the numbering scheme and image coordinates from \citet{lagattuta2017}.
The full table is available in machine-readable format in the online journal.}
\label{tab:images_a370}
\begin{tabular}{lccccc}
	\hline
	ID & RA ($^\circ$) & Dec ($^\circ$)  & $z$ & Rank & References \\
	\hline 
1.1 & $39.967083$ & $-1.576906$ & $0.804$ & \textsc{Gold} & 4 \\
1.2 & $39.976292$ & $-1.576042$ & $0.804$ & \textsc{Gold} & 4 \\
1.3 & $39.968683$ & $-1.576597$ & $0.804$ & \textsc{Gold} & 3,4 \\
2.1 & $39.973850$ & $-1.584225$ & $0.725$ & \textsc{Gold} & 1,3,4 \\
2.2 & $39.970954$ & $-1.585047$ & $0.725$ & \textsc{Gold} & 1,4 \\
2.3 & $39.968746$ & $-1.584519$ & $0.725$ & \textsc{Gold} & 1,4 \\
2.4 & $39.969425$ & $-1.584733$ & $0.725$ & \textsc{Gold} & 1,4 \\
2.5 & $39.969646$ & $-1.584842$ & $0.725$ & \textsc{Gold} & 1,4 \\
3.1 & $39.965650$ & $-1.566856$ & $1.955$ & \textsc{Gold} & 3,5 \\
3.2 & $39.968529$ & $-1.565811$ & $1.955$ & \textsc{Gold} & 3,5 \\
4.1 & $39.979650$ & $-1.576386$ & $1.273$ & \textsc{Gold} & 2,3,4 \\
4.2 & $39.970721$ & $-1.576269$ & $1.273$ & \textsc{Gold} & 3,4 \\
4.3 & $39.961937$ & $-1.577936$ & $1.273$ & \textsc{Gold} & 3,5 \\
6.1 & $39.969425$ & $-1.577206$ & $1.063$ & \textsc{Gold} & 4 \\
6.2 & $39.964329$ & $-1.578231$ & $1.063$ & \textsc{Gold} & 4 \\
6.3 & $39.979646$ & $-1.577092$ & $1.063$ & \textsc{Gold} & 4 \\
7.1 & $39.969775$ & $-1.580431$ & $2.751$ & \textsc{Gold} & 4 \\
7.2 & $39.969871$ & $-1.580772$ & $2.751$ & \textsc{Gold} & 4 \\
7.3 & $39.968808$ & $-1.585633$ & $2.751$ & \textsc{Gold}$^a$ & 5 \\
7.4 & $39.986554$ & $-1.577581$ & $2.751$ & \textsc{Gold}$^a$ & 5 \\
7.5 & $39.961542$ & $-1.580006$ & $2.751$ & \textsc{Gold}$^a$ & 5 \\
9.1 & $39.962400$ & $-1.577886$ & $1.518$ & \textsc{Gold} & 3,5\\
9.2 & $39.969483$ & $-1.576267$ & $1.518$ & \textsc{Gold} & 3 \\
9.3 & $39.982017$ & $-1.576533$ & $1.518$ & \textsc{Gold} & 3 \\
10.1 & $39.968567$ & $-1.571761$ & $2.751$ & \textsc{Gold} & 5 \\
10.2 & $39.968004$ & $-1.570875$ & $2.751$ & \textsc{Gold} & 5 \\
12.1 & $39.969596$ & $-1.566625$ & $3.481$ & \textsc{Gold} & 5 \\
12.2 & $39.959221$ & $-1.575244$ & $3.481$ & \textsc{Gold} & 5 \\
12.3 & $39.984117$ & $-1.570858$ & $3.481$ & \textsc{Gold}$^a$ & 5 \\
13.1 & $39.979537$ & $-1.571789$ & $4.247$ & \textsc{Gold} & 5 \\
13.2 & $39.975179$ & $-1.568825$ & $4.247$ & \textsc{Gold} & 5 \\
13.3 & $39.956753$ & $-1.577506$ & $4.247$ & \textsc{Gold} & 5 \\
14.1 & $39.972283$ & $-1.577983$ & $3.128$ & \textsc{Gold} & 4 \\
14.2 & $39.972192$ & $-1.580103$ & $3.128$ & \textsc{Gold} & 4 \\
14.3 & $39.974183$ & $-1.585608$ & $3.128$ & \textsc{Gold} & 4 \\
14.4 & $39.981313$ & $-1.578158$ & $3.128$ & \textsc{Gold} & 5 \\
14.5 & $39.957671$ & $-1.580447$ & $3.128$ & \textsc{Gold} & 5 \\
15.1 & $39.971328$ & $-1.580604$ & $3.708$ & \textsc{Gold} & 4 \\
15.2 & $39.971935$ & $-1.587051$ & $3.708$ & \textsc{Gold} & 4 \\
15.3 & $39.971027$ & $-1.577791$ & $3.708$ & \textsc{Gold} & 4 \\
15.4 & $39.984017$ & $-1.578451$ & $3.708$ & \textsc{Gold} & 5 \\
16.1 & $39.964016$ & $-1.588078$ & $3.774$ & \textsc{Gold} & 4 \\
16.3 & $39.984414$ & $-1.584111$ & $3.774$ & \textsc{Gold} & 5 \\
\end{tabular}
\end{table}

\begin{table}
\contcaption{Lensed Images for Abell 370.}
\label{tab:images_a370_cont}
\centering
\begin{tabular}{lcccccc}
	\hline
	ID & RA ($^\circ$) & Dec ($^\circ$)  & $z$ & Rank & References \\
	\hline
17.1 & $39.969758$ & $-1.588533$ & $4.257$ & \textsc{Gold} & 4 \\
17.2 & $39.985403$ & $-1.580841$ & $4.257$ & \textsc{Gold} & 5 \\
17.3 & $39.960235$ & $-1.583651$ & $4.257$ & \textsc{Gold} & 5 \\
18.1 & $39.975830$ & $-1.587061$ & $4.430$ & \textsc{Gold} & 4 \\
18.2 & $39.981476$ & $-1.582073$ & $4.430$ & \textsc{Gold} & 5 \\
18.3 & $39.957362$ & $-1.582086$ & $4.430$ & \textsc{Gold} & 5 \\
19.1 & $39.971996$ & $-1.587865$ & $5.649$ & \textsc{Gold} & 4 \\
19.2 & $39.985142$ & $-1.579094$ & $5.649$ & \textsc{Gold} & 5 \\
19.3 & $39.958316$ & $-1.581309$ & $5.649$ & \textsc{Gold} & 5 \\
20.1 & $39.965271$ & $-1.587803$ & $5.750$ & \textsc{Gold} & 4 \\
20.2 & $39.963608$ & $-1.586883$ & $5.750$ & \textsc{Gold} & 4 \\
22.1 & $39.974408$ & $-1.586100$ & $3.128$ & \textsc{Gold} & 4 \\
22.2 & $39.981675$ & $-1.579686$ & $3.128$ & \textsc{Gold} & 5 \\
22.3 & $39.957906$ & $-1.581011$ & $3.128$ & \textsc{Silver}$^a$ & 5 \\
23.1 & $39.980254$ & $-1.566764$ & $5.939$ & \textsc{Gold} & 5 \\
23.2 & $39.957314$ & $-1.572744$ & $5.939$ & \textsc{Gold} & 5 \\
23.3 & $39.977165$ & $-1.566275$ & $5.939$ & \textsc{Gold} & 5 \\
24.1 & $39.963113$ & $-1.570594$ & $4.915$ & \textsc{Gold} & 5 \\
24.2 & $39.962029$ & $-1.572336$ & $4.915$ & \textsc{Gold} & 5 \\
25.1 & $39.987325$ & $-1.578867$ & $3.808$ & \textsc{Gold} & 5 \\
25.2 & $39.961950$ & $-1.583169$ & $3.808$ & \textsc{Gold} & 5 \\
26.1 & $39.979939$ & $-1.571390$ & $3.936$ & \textsc{Gold} & 5 \\
26.2 & $39.974464$ & $-1.568094$ & $3.936$ & \textsc{Gold} & 5 \\
26.3 & $39.957165$ & $-1.576958$ & $3.936$ & \textsc{Gold} & 5 \\
27.1 & $39.980692$ & $-1.571117$ & $3.016$ & \textsc{Gold} & 5 \\
27.2 & $39.958290$ & $-1.575907$ & $3.016$ & \textsc{Gold} & 5 \\
27.3 & $39.972442$ & $-1.567161$ & $3.016$ & \textsc{Gold} & 5 \\
28.1 & $39.963492$ & $-1.582281$ & $2.911$ & \textsc{Gold} & 5 \\
28.2 & $39.967058$ & $-1.584558$ & $2.911$ & \textsc{Gold} & 5 \\
28.3 & $39.987816$ & $-1.577453$ & $2.911$ & \textsc{Gold} & 5 \\
30.1 & $39.983459$ & $-1.570449$ & $5.646$ & \textsc{Gold} & 5 \\
30.2 & $39.972404$ & $-1.566353$ & $5.646$ & \textsc{Gold} & 5 \\
32.1 & $39.966286$ & $-1.569345$ & $4.495$ & \textsc{Gold} & 5 \\
32.2 & $39.988098$ & $-1.575187$ & $4.495$ & \textsc{Gold} & 5 \\
32.3 & $39.960682$ & $-1.578380$ & $4.495$ & \textsc{Gold} & 5 \\
33.1 & $39.962723$ & $-1.586003$ & $4.882$ & \textsc{Gold} & 5 \\
33.2 & $39.966217$ & $-1.587996$ & $4.882$ & \textsc{Gold} & 5 \\
34.1 & $39.970108$ & $-1.570150$ & $5.244$ & \textsc{Gold} & 5 \\
34.2 & $39.971806$ & $-1.588040$ & $5.244$ & \textsc{Gold} & 5 \\
34.3 & $39.958565$ & $-1.581701$ & $5.244$ & \textsc{Gold} & 5 \\
34.4 & $39.985046$ & $-1.579559$ & $5.244$ & \textsc{Gold} & 5 \\
35.1 & $39.981541$ & $-1.565862$ & $6.173$ & \textsc{Gold} & 5 \\
35.2 & $39.975826$ & $-1.564442$ & $6.173$ & \textsc{Gold} & 5 \\
36.1 & $39.962444$ & $-1.580710$ & $6.285$ & \textsc{Gold} & 5 \\
36.2 & $39.965996$ & $-1.584384$ & $6.285$ & \textsc{Gold} & 5 \\
38.1 & $39.977154$ & $-1.573792$ & $3.156$ & \textsc{Gold} & 5 \\
38.2 & $39.975071$ & $-1.572116$ & $3.156$ & \textsc{Gold} & 5 \\
\hline
\end{tabular}
\end{table}

%%%%%%%%%%%%%%%%%%%%%%%%%%%%%%%%%%%%%%%%%%%%%%%%%%

% Don't change these lines
\bsp	% typesetting comment
\label{lastpage}
\end{document}